\documentclass[twocolumn]{aastex631}

\usepackage{graphicx}
\usepackage{gensymb}

\newcommand{\loghone}{log $N_\mathrm{HI}$}
\newcommand{\logh}{log $n_\mathrm{H}$}
\newcommand{\vlsr}{$v_\mathrm{LSR}$}
\newcommand{\kms}{km\,s$^{-1}$}

\newcommand{\cm}{cm$^{-2}$}
\newcommand{\hd}{\object{HD\,156359}}
\newcommand{\hi}{\ion{H}{1}}
\newcommand{\oi}{\ion{O}{1}}
\newcommand{\cloudy}{\textsc{Cloudy}}

\received{September 28, 2022}

\submitjournal{ApJ}

\shorttitle{Caught in the Act: A Metal-Rich High-Velocity Cloud in the Inner Galaxy}
\shortauthors{Cashman, Fox, Wakker et al.}

\begin{document}
\title{Caught in the Act: A Metal-Rich High-Velocity Cloud in the Inner Galaxy}

\correspondingauthor{Frances Cashman, Andrew Fox}
\email{frcashman@stsci.edu, afox@stsci.edu}

\author[0000-0003-4237-3553]{Frances H. Cashman}
\affiliation{Space Telescope Science Institute, 3700 San Martin Drive, Baltimore, MD 21218, USA}

\author[0000-0003-0724-4115]{Andrew J. Fox}
\affiliation{AURA for ESA, Space Telescope Science Institute, 3700 San Martin Drive, Baltimore, MD 21218, USA}

\author[0000-0002-0507-7096]{Bart P. Wakker}
\affiliation{Department of Astronomy, University of Wisconsin-Madison, 475 North Charter Street, Madison, WI 53706, USA}

\author[0000-0002-6541-869X]{Trisha Ashley}
\affiliation{Space Telescope Science Institute, 3700 San Martin Drive, Baltimore, MD 21218, USA}


\author[0000-0002-9139-2964]{Derck Massa}
\affiliation{Space Science Institute, 4750 Walnut Street Suite 205 Boulder, Colorado 80301, USA}

\author[0000-0003-1892-4423]{Edward B. Jenkins}
\affiliation{Department of Astrophysical Sciences, Princeton University, Princeton, NJ 08544-1001, USA}

\author[0000-0002-7955-7359]{Dhanesh Krishnarao}
\affiliation{NSF Astronomy \& Astrophysics Postdoctoral Fellow, Johns Hopkins University, 3400 N. Charles Street, Baltimore, MD 21218, USA}
\affiliation{Department of Physics, Colorado College, 14 East Cache La Poudre Street, Colorado Springs, CO 80903, USA}
\affiliation{Space Telescope Science Institute, 3700 San Martin Drive, Baltimore, MD 21218, USA}

\author[0000-0002-6050-2008]{Felix J. Lockman}
\affiliation{Green Bank Observatory, P.O. Box 2, Rt. 28/92, Green Bank, WV 24944, USA}

\author[0000-0002-8109-2642]{Robert A. Benjamin}
\affiliation{Department of Physics, University of Wisconsin-Whitewater, 800 West Main Street, Whitewater, WI 53190, USA}

\author[0000-0002-3120-7173]{Rongmon Bordoloi}
\affiliation{Department of Physics, North Carolina State University, 421 Riddick Hall, Raleigh, NC 27695-8202, USA}

\author{Tae-Sun Kim}
\affiliation{Department of Astronomy, University of Wisconsin-Madison, 475 North Charter Street, Madison, WI 53706, USA}



\begin{abstract}
We characterize the chemical and physical conditions in an outflowing high-velocity cloud in the inner Galaxy. 
We report a super-solar metallicity of [O/H] = $+0.36\pm0.12$ for the high-velocity cloud at \vlsr\ = 125.6 \kms\ toward the star \hd\ ($l$ = 328.$\degree$7, $b$ = $-$14.$\degree$5, $d$ = 9 kpc, $z$ = $-$2.3 kpc). 
Using archival observations from FUSE, HST STIS, and ESO FEROS we measure high-velocity absorption in \ion{H}{1}, \oi, \ion{C}{2}, \ion{N}{2}, \ion{Si}{2}, \ion{Ca}{2}, \ion{Si}{3}, \ion{Fe}{3}, \ion{C}{4}, \ion{Si}{4}, \ion{N}{5}, and \ion{O}{6}. 
We measure a low \hi\ column density of log $N$(\hi) = $15.54\pm0.05$ in the HVC from multiple unsaturated \hi\ Lyman series lines in the FUSE data. 
We determine a low dust depletion level in the HVC from the relative strength of silicon, iron, and calcium absorption relative to oxygen, with [Si/O]=$-0.33\pm0.14$, [Fe/O]=$-0.30\pm0.20$, and [Ca/O] =$-0.56\pm0.16$. 
Analysis of the high-ion absorption using collisional ionization models indicates that the hot plasma is multi-phase, with the \ion{C}{4} and \ion{Si}{4} tracing 10$^{4.9}$ K gas and \ion{N}{5} and \ion{O}{6} tracing 10$^{5.4}$ K gas. 
The cloud's metallicity, dust content, kinematics, and close proximity to the disk are all consistent with a Galactic wind origin. 
As the \hd\ line of sight probes the inner Galaxy, the HVC appears to be a young cloud caught in the act of being entrained in a multi-phase Galactic outflow and driven out into the halo.
\end{abstract}




\section{Introduction} \label{sec:intro}

\begin{figure*}[!ht]
    \centering
    \includegraphics[scale=0.43]{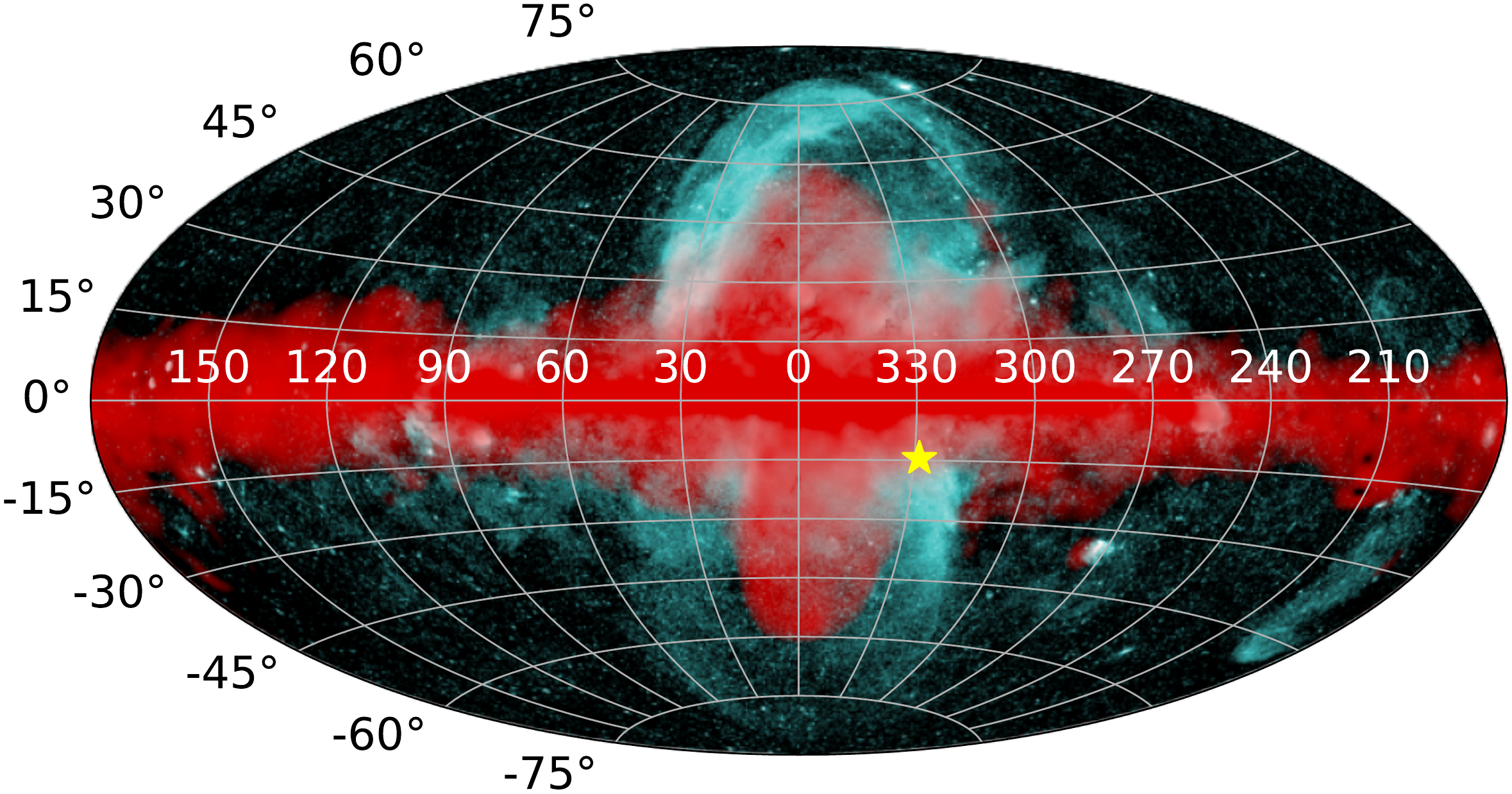}
    \includegraphics[scale=0.13]{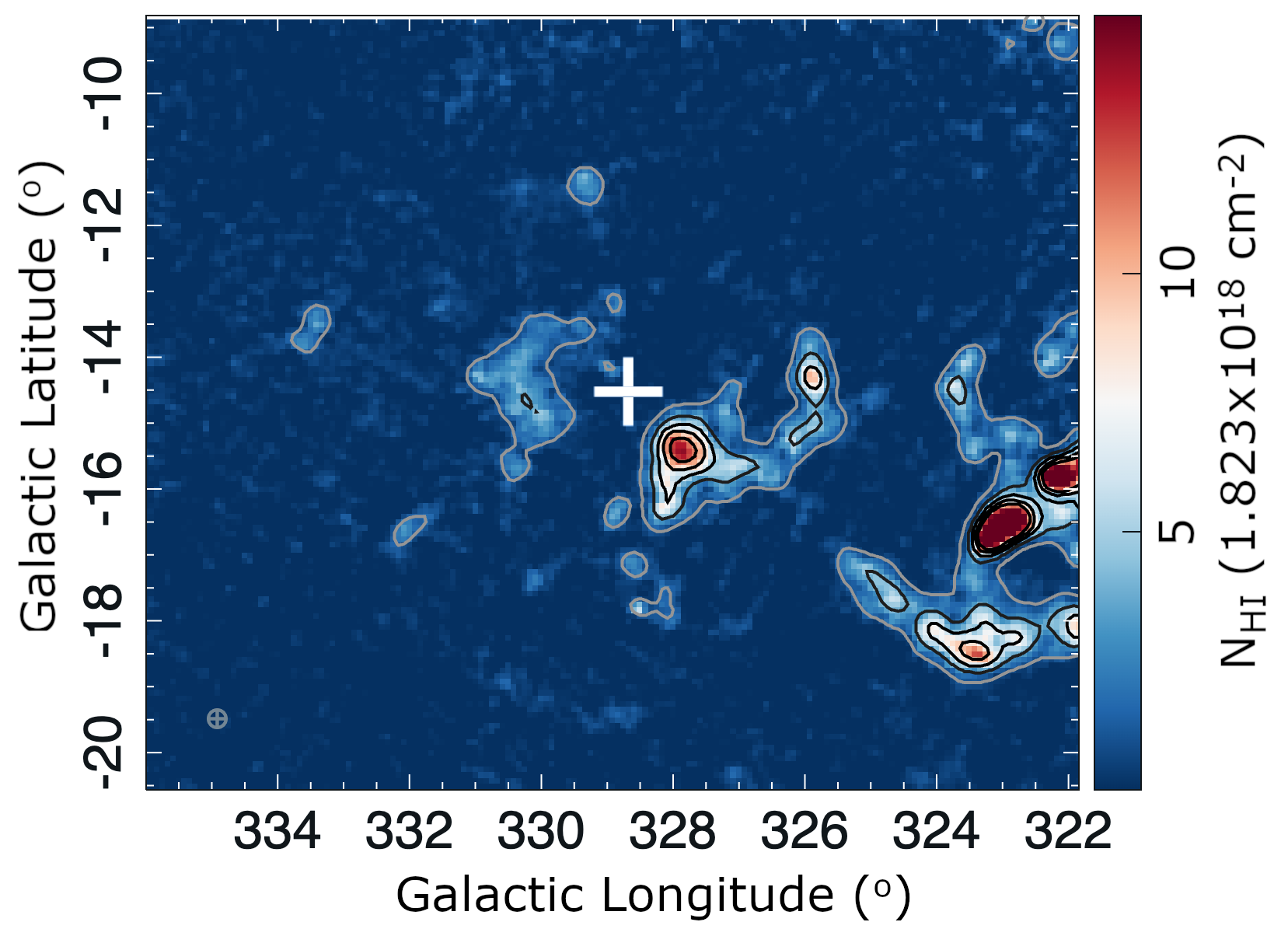}
    \caption{Left panel: The location of \hd\ with reference to the Fermi and eROSITA bubbles. The composite Fermi–eROSITA image from \citet{predehl2020}, where the softer X-ray emission (0.6--1 keV, in cyan) envelopes the harder component of the extended GeV emission of the Fermi bubbles (in red; adapted from \citealt{selig2015}). Right panel: \hi\ column density map from the 21\,cm HI4PI survey showing the distribution of \ion{H}{1} in the region from 100--150 \kms\ \citet{hi4pi2016} in a section of ``Complex WE'' by \citet{wakker1991}. The location of \hd\ is marked with a white cross. The contours correspond to \loghone\ = 2.08, 6.23, 10.4, 14.5, 18.7 $\times$ 10$^{18}$ \cm.} 
    \label{fig:allsky}
\end{figure*}

The supermassive black hole, Sagittarius A$^*$, and surrounding regions of active star formation power an outflowing multiphase wind from the center of the Milky Way (MW). Outflows play a critical role in the baryon cycle, the cycling of gas into and out of galaxies, which helps regulate the evolution of galaxies. The MW offers us a front-row seat view to study outflows over a range of wavelengths and phases and understand their impact on galaxy evolution.

Presently, the farthest-reaching known outflows in the MW are the Fermi \citep{su2010,ackermann2014} and eROSITA \citep{bland2003,predehl2020} Bubbles, which are giant gamma- and X-ray emission extending $\sim$10 kpc and 14 kpc above and below the Galactic disk, respectively (see Figure \ref{fig:allsky}). The origin of the Fermi Bubbles is believed to be an energetic GC Seyfert flare event $\sim$3.5 Myr ago \citep{bland2019}. 

Ultraviolet (UV) absorption-line studies have found gas outflowing at high-velocity 
in a number of sight lines passing through or near the Fermi Bubbles (see \citealt{keeney2006,zech2008,fox2015,bordoloi2017,savage2017,karim2018,ashley2020, ashley2022}). 
These high-velocity clouds (HVCs) consist of neutral (e.g., \ion{C}{1}, \ion{O}{1}), singly ionized (\ion{S}{2}, \ion{Si}{2}, \ion{Fe}{2}), and highly ionized (\ion{C}{4}, \ion{Si}{4}, \ion{O}{6}) gas. Cold gas has also been detected in the nuclear outflow via emission in neutral hydrogen and molecules. Several hundred \hi\ 21 cm clouds have been detected at low latitudes in the Fermi Bubbles outflowing from the Galactic Center (GC) \citep{mcclure2013,diteodoro2018,lockman2020}. Carbon monoxide (CO) has been detected in submillimeter emission in two dense molecular clouds comoving within the \hi\ 21 cm outflow \citep{diteodoro2020}. More recently, molecular hydrogen was discovered in the Galactic nuclear outflow $\sim$1 kpc below the GC \citep{cashman2021}. Outflowing gas associated with the Fermi Bubbles has also been seen in optical emission \citep{krishnarao2020b}. Together, these observations of HVCs near the Galactic Center provide observational constraints on the properties of the MW nuclear wind.

The outflowing wind interacts with gas in the disk in multiple ways.
First, it can disrupt the position and velocity of the disk \citep{krishnarao2020a}, leading to warps and perturbations.
Second, it can accelerate and entrain cool gas into the halo. In some instances, this high-velocity gas may cool, lose buoyancy, and fall back onto the disk as a ``Galactic fountain", supplying new fuel for further star formation (see \citealt{shapiro1976,bregman1980,kahn1981,deavillez1999}). In other cases, the high-velocity gas may survive being accelerated into the halo, and even escape. Cloud survival is the subject of recent theoretical work focusing on how cool gas clouds develop and grow in the hot wind (see \citealt{gronke2020, sparre2020}). 

Metallicities of HVCs are important to determine because they provide direct evidence for the origin of the clouds. Fully UV-based metallicities (with the metal and \hi\ column densities measured along the same UV sight lines) are ideal to ensure that the metal absorption and hydrogen absorption are tracing the same gas, but they are rare because of the difficulty in isolating unsaturated high-velocity \hi\ components in UV absorption. 
These high-velocity \hi\ components are often saturated or blended, even for high-order Lyman series lines, thus there is only a narrow range of $N$(\hi) values for which precise measurements are possible \citep{french2021}.
Therefore HVC metallicities are frequently determined from a combination of a UV measurement along an infinitesimal sight line with a \hi\ measurement made using a much larger 21 cm beam. However, this combination introduces beam-smearing uncertainties on the metallicity. Currently, the only fully-UV-based metallicities are for two HVCs toward M5-ZNG1 \citep{zech2008}, but this is a high-latitude direction ($b\approx50\degree$) far from the GC. 

At lower latitudes, we can target UV-bright stars as background sources, but very few UV spectra of distant GC stars ($d\gtrsim8$\,kpc) exist. Therefore, it is crucial to fully analyze the few spectra available, including LS 4825 \citep{savage2017, cashman2021} and HD 156359 (this paper).
It is precisely these low-latitude sight lines that are likely to harbor young clouds that have only recently been entrained in the Galactic outflow. Finding an outflowing cloud at low latitude means catching it close to its origin, thus offering a rare opportunity to study the cloud before it has undergone significant mixing as it begins its journey into the halo. Observing recently entrained clouds therefore gives a snapshot of outflowing gas when it first exits the disk.

In this paper we present a detailed spectroscopic analysis of 
an HVC detected toward an inner Galaxy sight line, \hd.
By studying the properties of this cloud, we characterize the physical and chemical conditions of the gas in the inner Galaxy, in a region likely influenced by the nuclear wind.

\section{Data} \label{sec:data}

\subsection{The HD 156359 Sight line}\label{subsec:hd156359}
HD 156359 (at $l,b,d$ = 328.68\degree, $-$14.52\degree, 9 kpc) is one of the best-studied GC sight lines \citep{sembach1991,sembach1995}.  
The sight line lies in the inner Galaxy at the boundary of the southern Fermi Bubble and within the eROSITA X-ray bubble \citep{predehl2020}, a region of enhanced X-ray emission (see Figure \ref{fig:allsky}). This sight line also passes close to a complex of small HVCs dubbed ``Complex WE'' by \citet{wakker1991}, less than 1$\degree$ from one of the densest cores.
\citet{sembach1991} classified \hd\ as a O9.7 Ib--II star on the basis of stellar photospheric lines and wind profiles in its UV spectrum. The spectral type and apparent magnitude of the star yield a spectroscopic distance of 11.1$\pm$2.8 kpc. 
However, recent Gaia EDR3 parallax measurements \citep{bailer2021} place \hd\ at a distance of 9.0$^{+3.1}_{-2.1}$ kpc, which implies a $z$-distance below the plane of 2.3 kpc. We adopt the Gaia distance in our analysis.

The sight line toward \hd\ intersects at least three spiral arms, the 
Sagittarius, Scutum, and Norma arms. Spiral-arm models from \citet{vallee2017} give 
the expected velocities of the Sagittarius, Scutum, and Norma arms at approximately 
$-$10, $-$55, and $-$103 \kms, respectively. Thus, absorption components observed at 
these velocities can be attributed to gas in (or associated with) the spiral arms. 
However, the spiral arms cannot explain the HVC observed at $v_{\rm LSR}$=125~\kms.

High-ion absorption toward \hd\ was first observed with the International Ultraviolet Explorer 
\citep[IUE;][]{sembach1991} and later by the Goddard High Resolution Spectrograph \citep[GHRS;][]{sembach1995}. 
The high-ion information in the FUSE and HST STIS spectra of this sight line has not been previously published 
-- the sight line is not included in the FUSE \ion{O}{6} survey of Galactic disk sight lines by \citet{bowen2008}, as that survey was restricted to $|b|<10\degree$. We also include analysis of a single archival optical spectrum of \hd\ taken with FEROS at the European Southern Observatory (ESO) at La Silla. Details of all observations of spectra used in this paper are described below.

\subsection{UV Observations}\label{subsec:uv_obs}

\hd\ was observed by FUSE \citep{moos2000}
under programs P101, S701, and U109 between 2000 April 12 and 2006 April 25 (PIs Sembach, Andersson, and Blair, respectively). The raw spectra were downloaded from the MAST FUSE archive, and the \texttt{CalFUSE} pipeline (v3.2.2; \citealt{dixon2009}) was used to reduce and extract the spectra. Data from the SiC, LiF1, and LiF2 channels were used for the analysis. A detailed explanation of the data reduction, as well as refinements to the \texttt{CalFUSE} data reduction procedures can be found in \citet{wakker2003} and \citet{wakker2006}.
The spectra have a signal-to-noise ratio (S/N) $\sim$12--26 per resolution element and a velocity resolution of 20 km s$^{-1}$ (FWHM). The data were binned by three pixels for the fitting analysis. The FUSE wavelength coverage is $\sim$912--1180{\AA}.

A single exposure of HD 156359 was obtained with HST/STIS on 2003 March 26 under program 9434 (PI Lauroesch) using the E140M grating. The data were downloaded from the MAST HST archive and reduced using \texttt{calstis} (v.3.4.2, \citealt{dressel2007}). 
The echelle orders were combined to create a single continuous spectrum, and in regions of order-overlap spectral counts were combined to increase the S/N ratio.
The data have S/N $\sim$10--38 per resolution element and a FWHM velocity resolution of 6.5 km s$^{-1}$, i.e. spectral resolution ($\lambda$/$\Delta\lambda$) $\sim$45,800. The STIS E140M wavelength coverage is $\sim$1160--1725{\AA}. 
Wavelengths and velocities for the absorption-line features are given in the local standard of rest (LSR) reference frame, where the correction factor \vlsr\ $-$ $v_\mathrm{Helio}$ = $-0.35$ \kms\ for the direction toward \hd. 

\subsection{Optical Observations}\label{subsec:opt_obs}

A spectrum of \hd\ was captured using the Fibre-fed Extended Range Optical Spectrograph (FEROS) at the European Southern Observatory (ESO) La Silla 2.2 meter telescope on 2006 April 30 (PI: Bouret, PID: 077.D-0635). The retrieved archival product covers the wavelength range $\sim$3565--9214 {\AA} with R = 48000 and FWHM = 6.25 \kms. The primary data product was reduced automatically using the FEROS Data Reduction Software (DRS) pipeline version \texttt{fern/1.0}. In the automated reduction process, the bias is subtracted using overscan regions and bad columns are replaced by the mean values of the neighboring columns. The orders are rectified and then extracted using the standard method. Next, the extracted spectra are flat-fielded and wavelength calibrated, then rebinned to a constant dispersion of 0.03 {\AA}. Finally, the individual orders are combined into a single 1D spectrum. As the archival reduced data are not flux calibrated, the continuum of the stellar spectrum was normalized using the \texttt{linetools} software package.

\section{Measurements}\label{sec:meas}

In this section we describe our absorption-line measurement processes, including stellar continuum fitting 
and procedures for measuring lines of different ionization states.
Our measurements are based on the \textsc{VPFIT} software program (v12.2; \citealt{carswell2014}),
which we used to conduct a set of Voigt-profile fits to the absorption-line profiles, with wavelengths and oscillator strengths taken from the compilations of \citet{morton2003} and \citet{cashman2017}. The measured lines are listed in Table \ref{tab:fdata}. 

\subsection{Stellar continuum modeling}\label{subsec:tlusty}

Given the spectral type of \hd\ \citep[O9.7 Ib--II;][]{sembach1991}, the stellar continuum placement requires careful consideration. 
In addition to estimating the stellar continuum through comparison with FUSE spectra of stars of similar spectral type, we also constructed a TLUSTY model \citep{lanz2003} of this type of star as a continuum placement guide. This was particularly useful in regions where the interstellar absorption was stronger.  

\begin{figure}
    \centering
    \includegraphics[scale=0.27]{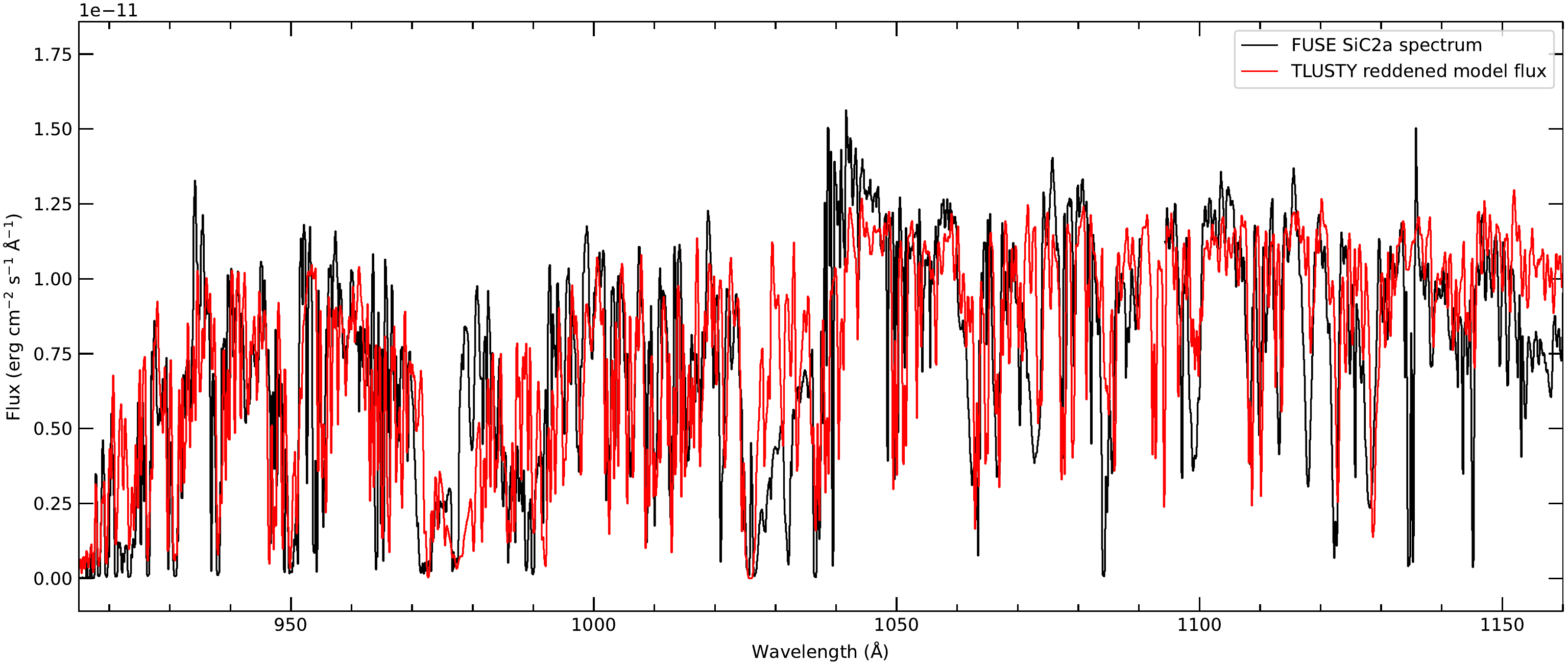}
    \caption{A comparison of a section of the FUSE FUV spectrum with a TLUSTY stellar model for the same spectral type, O9.7, which aided in the continuum placement. The FUSE SiC2a spectrum from 914--944 {\AA} is shown in black with a stellar TLUSTY model shown in red. The TLUSTY model has $T_\mathrm{eff}$ = 26000 K, log $g$ = 3.0, $v_\mathrm{turb}$ = 10.0 km s$^{-1}$, and is rotationally broadened by 90 km s$^{-1}$ in order to match the observed \ion{Si}{3} 1300 {\AA} triplets. The model is reddened by an $E(B-V)$ = 0.09 using the mean \citet{fitzpatrick2007} extinction curve and Lyman line absorption for log $N$(\hi) = 20.78 \citep{sembach1991} is also applied. The flux was scaled by 1.25$\times$10$^{-20}$ to provide an overall agreement with observations to $\sim$10\%. 
    }
    \label{fig:app_tlusty}
\end{figure}

The adopted TLUSTY model has $T_\mathrm{eff}$ = 26000 K, log $g$ = 3.0, $v_\mathrm{turb}$ = 10.0 \kms\ and is rotationally broadened by 90 \kms\ in order to match the \ion{Si}{3} 1300 {\AA} triplets. We matched the observed UV wind signatures found in the observed FUSE and STIS spectra, where all observed spectra are adjusted for the stellar radial velocity $v_\mathrm{rad} = -82$ \kms\ \citep{gontcharov2006}. The model is reddened by an $E(B-V) = 0.09$, using the mean \citet{fitzpatrick2007} extinction curve (extrapolated to 912 {\AA}), and Lyman series absorption for a foreground \hi\ column of $N$(\hi) = 6.0 $\times$ 10$^{20}$ \cm\ is also included \citep{sembach1991}. The model is then scaled by 1.25 $\times$ 10$^{-20}$ in order to match observations to $\sim$10\%. Finally, all spectra were binned to 0.1 {\AA}, $\sim$15 to 30 \kms, depending on wavelength, see Figure \ref{fig:app_tlusty}. When consulting the model as a guide for interstellar line continua, additional tweaks of about 5\% were needed to match local continua. 
Our subsequent column density measurements account for the uncertainty inherent in the continuum-placement process.

\subsection{\hi\ absorption}\label{subsec:hi-meas}

\begin{figure*}[!ht]
    \centering
    \includegraphics[scale=0.41]{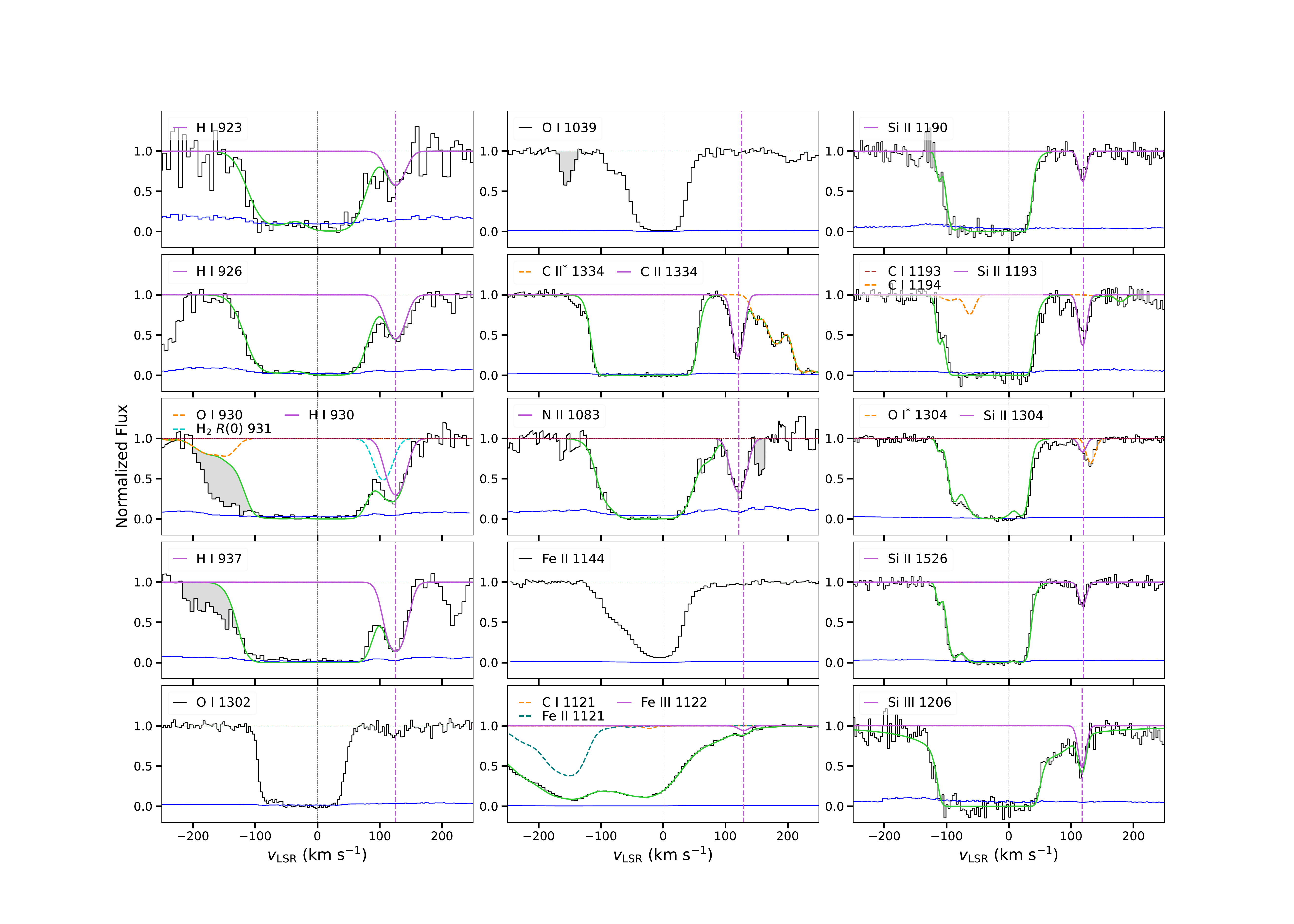}
    \caption{Velocity profiles of the neutral, low-, and intermediate-ion absorption lines detected toward HD 156359. The normalized flux is shown in black, the continuum level is in red, and the 1$\sigma$ error in the normalized flux is in blue. The vertical line at 0 km s$^{-1}$ marks the region associated with the Milky Way. Each of these profiles shows a Voigt fit to the data except for \oi\, for which we provide an AOD measurement, and for \ion{Fe}{2}, which is a non-detection. The solid green curve is the overall Voigt profile fit to the absorption. The magenta curve is the Voigt profile fit to the HVC absorption feature near $+$125 \kms\ and the vertical dashed magenta line marks the velocity centroid of the fitted component. }
    \label{fig:vplow}
\end{figure*}

\begin{figure}[!ht]
    \centering
    \includegraphics[scale=0.4]{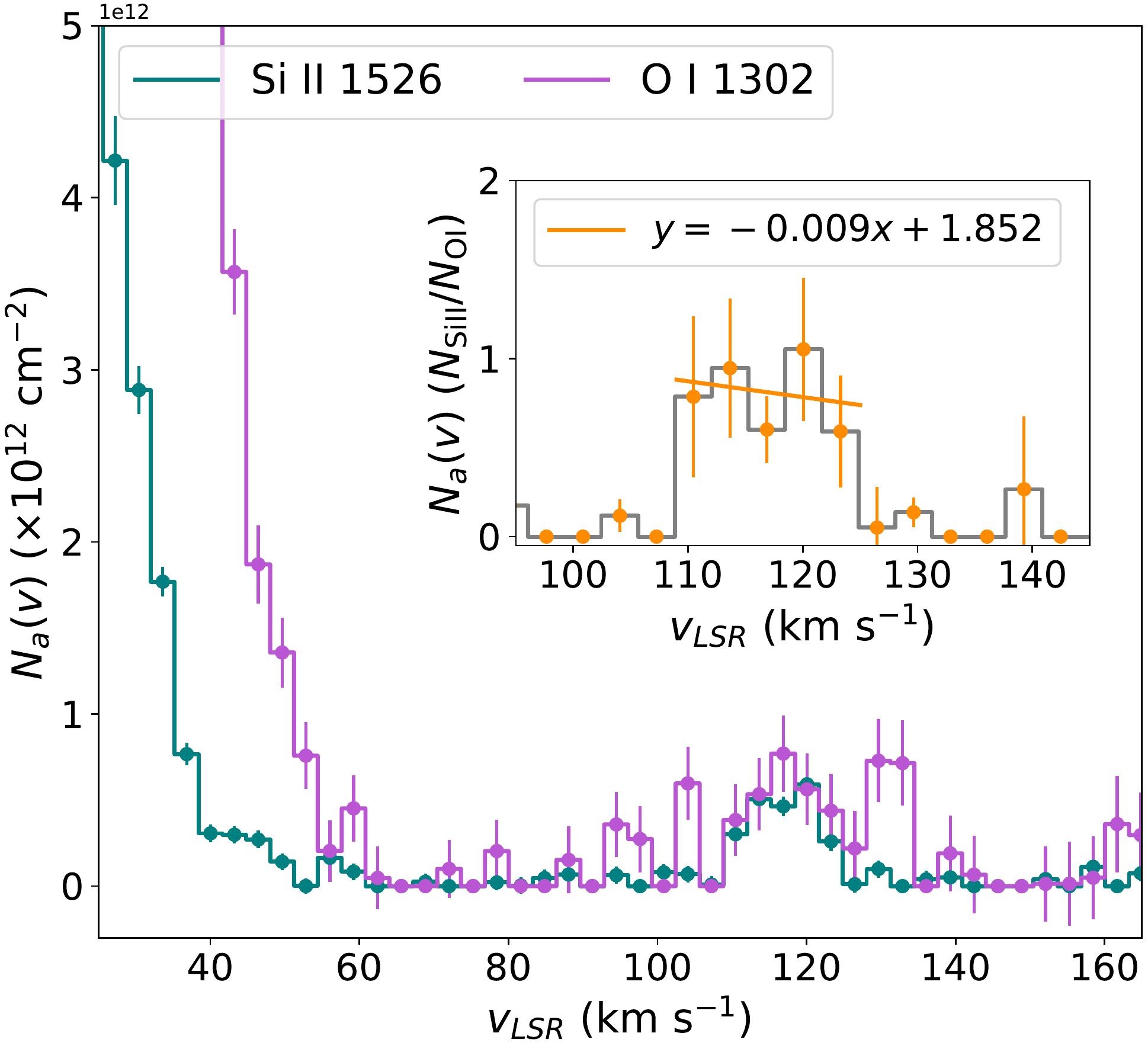}
    \caption{A comparison of the normalized apparent column density plots between \oi\ 1302 in violet and \ion{Si}{2} 1526 in green. The profiles have similar shapes although the \oi\ 1302 profile shows a lower signal to noise. The inset panel shows the apparent column density ratio of \ion{Si}{2} 1526 to \oi\ 1302. A linear fit to 5 pixels in the region from $+110 \lesssim v \lesssim 125$~\kms\ has a slope near zero within the margin of error, supporting the reality of the detection in both ions. 
    }
    \label{fig:nav}
\end{figure}

\begin{figure}
    \centering
    \includegraphics[scale=0.48]{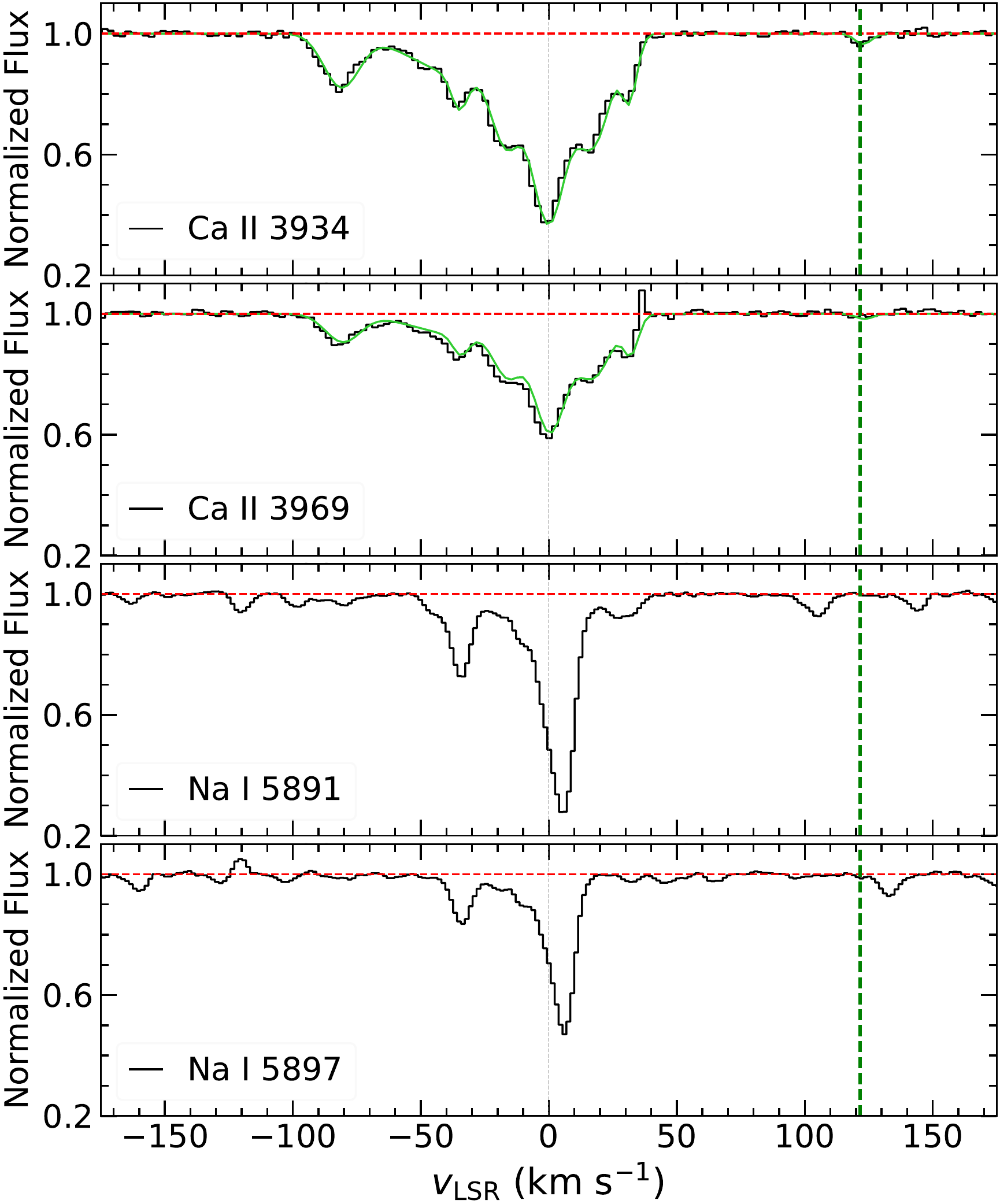}
    \caption{Velocity profiles of the \ion{Ca}{2} $\lambda\lambda$3934, 3969 and \ion{Na}{1} $\lambda\lambda$5891, 5897 absorption lines detected toward HD 156359 in the archival FEROS spectrum. The normalized flux is shown in black, the continuum level is in red. The gray vertical line at 0 km s$^{-1}$ marks the absorption region associated with the Milky Way. The green curves in the top two panels are the Voigt profile fits to the data, where the green vertical line at 123.7 \kms\ marks the location of the HVC. There is a non-detection of high-velocity \ion{Na}{1}, although telluric H$_2$O absorption lines are present in this region. The lower end of the y-axis begins at $+0.2$ in all panels to improve the visibility of the absorption features.
    }
    \label{fig:RS}
\end{figure}

The FUSE spectrum of \hd\ covers almost the entire \ion{H}{1} Lyman series, from Lyman-$\beta$ at 1025 \AA\ down to the Lyman limit at 912 \AA. However, we limited our fitting to \ion{H}{1} lines in regions where the stellar continuum was better-defined and showed the least contamination from stellar features and/or interstellar molecular lines, which are prolific in the FUSE bandpass. 

We selected \hi\ $\lambda\lambda$923, 926, 930, and 937 (see Figure \ref{fig:vplow}) to derive the \hi\ column density. We fit a continuum to each of these lines locally. In order to account for sensitivity to continuum placement, we consider a high and low continuum placement for our column density measurement across the wavelength range $\lambda$917--944. \hi\ $\lambda$923 lies in a noisy region of the upper Lyman lines frequently associated with the Inglis-Teller effect \citep{inglis1939}, in which overlapping stellar absorption lines have merged to produce the appearance of a depressed continuum. 
We perform a simultaneous Voigt profile fit to $\lambda\lambda$926, 930, 937 with the positions of the low-velocity absorption initially based on weak ISM lines. We derive log $N_\mathrm{HI}$ = 15.54 $\pm$ 0.02 $\pm$ 0.05 in the HVC at $+125.6$ \kms, where the first error is the statistical error due to photon noise and the second is the systematic continuum-placement error. The result of these fits is shown Table \ref{tab:colden} and in Figure \ref{fig:vplow}, where we show that the simultaneous fit is also consistent with the noisier \hi\ $\lambda$923 profile. We include a detailed explanation of our continuum placement procedures and how we considered contamination from stellar features in Appendix \ref{sec:global}.

An important step in measuring \hi\ absorption in FUSE spectra is to decontaminate H$_2$ absorption.
Fortunately, the FUSE LiF1 and LiF2 channels provide us with an opportunity to model isolated H$_2$ lines. This allows us to account for the Milky Way's molecular contribution to the interstellar absorption lines blended with \hi\ in the SiC2A spectrum. We conducted an H$_2$ decontamination and describe and illustrate this process in detail in Appendix \ref{sec:h2-decontaminate}. 

\subsection{\oi\ absorption}\label{subsec:oi-meas}

After placing a smooth continuum through the \oi\ line at 1302 {\AA} in the STIS E140M spectrum, we noticed a small absorption feature near $+125$ \kms\ (see Figure \ref{fig:vplow}). This feature spans five pixels and has a significance of 3.2$\sigma$ ($EW$/$\sigma_{EW}$=3.2).
Since only one STIS exposure exists, 
we cannot confirm the \oi\ detection in another dataset,  
and this feature is not seen in the much weaker \oi\ line at 1039 {\AA} in the FUSE spectrum. 
To explore whether this feature is real, we compared the absorption profile to another low-ion line. 
Figure \ref{fig:nav} shows an apparent column density profile comparison of the \oi\ 1302 {\AA} line to the weak \ion{Si}{2} 1526 {\AA} line. We chose $\lambda$1526 because it is unsaturated, unblended, and seen in a region with high signal to noise. Although noisier, the profile of \oi\ $\lambda$1302 is very similar to \ion{Si}{2} $\lambda$1526 in the velocity region near $+125$ \kms, and the ratio of their apparent column densities is flat with velocity. To confirm this similarity, the inset plot in Figure \ref{fig:nav} shows a linear fit to the ratio over five pixels (over 2 resolution elements) with a slope equal to zero within the margin of error, as expected for a genuine \oi\ detection. 
Therefore, we proceed with treating the \oi\ $\lambda$1302 feature as real on the basis that: 
1) the line is detected at 3.2$\sigma$ significance,  
2) there is close kinematic consistency with \ion{Si}{2}, and 
3) the feature is centered at the same velocity as multiple other ions, e.g., \ion{H}{1}, \ion{C}{2}, \ion{N}{2}, and \ion{Ca}{2}. 
After applying a low and high continuum to this region, we measure log $N_\mathrm{OI}$ = 12.43 $\pm$ 0.08 $\pm$ 0.07, which includes the statistical error and a continuum-placement (systematic) error. This measurement is the foundation of our metallicity measurement in the HVC, which we discuss in Section \ref{sec:res_low}. 

\subsection{Low and intermediate ion absorption}

We detected \ion{C}{2} $\lambda$1334, \ion{Si}{2} $\lambda\lambda$1190, 1193, 1304, 1526, \ion{N}{2} $\lambda$1083, \ion{Si}{3} $\lambda$1206, and \ion{Fe}{3} $\lambda$1122 in the HVC near $+125$ \kms\ in the FUSE and STIS spectra (see Figure \ref{fig:vplow}). Our absorption-line measurements of these lines are given in Table \ref{tab:colden}. 
Achieving a simultaneous Voigt profile fit using all Si lines is hindered by difficulties in continuum placement for the stronger lines since large spans of the variable stellar continuum are absorbed. Instead, we adopt the Voigt profile measurement log $N_\mathrm{SiII}$ = 12.94$\pm$0.07 for the weakest unblended line available at 1526 {\AA}. We show that it is a good fit for $\lambda\lambda$1190, 1193, 1304 in Figure \ref{fig:vplow}.

Although detected, \ion{N}{2} 1083 lies in a portion of the FUSE spectrum where the stellar continuum is highly variable over a short range in wavelength. Comparing the neighboring low-velocity H$_2$ $J$3 1084 {\AA} line to other $J$3 lines in regions with smooth continua reveals that the continuum must drop significantly in this region, making it difficult to estimate log $N_\mathrm{NII}$. For this reason, our measurement for \ion{N}{2} is an upper limit. \ion{Fe}{2} 1144 has a low absorption profile in the vicinity of the HVC. We applied a high, middle, and low continuum across the absorbing region of $\lambda$1144 and calculate a significance of 2.2$\sigma$ after including a continuum placement error. We find a 3$\sigma$ limiting column density of log $N_\mathrm{FeII}$ $\leq$ 12.99, however, given the \ion{Fe}{3} detection in the HVC, we use the \ion{Fe}{3} column density in metallicity calculations going forward. In addition to the high-velocity absorption we observe for \ion{Fe}{3} $\lambda$1122, we detect an intermediate-velocity cloud (IVC) centered near $+$86 \kms\ with log $N$ = 12.98$\pm$0.13.

The archival ESO FEROS optical spectrum covers the \ion{Ca}{2} K and H and \ion{Na}{1} D lines, see Figure \ref{fig:RS}. Telluric H$_2$O absorption lines are present in the velocity range of the HVC, however, there is a non-detection of high-velocity \ion{Na}{1} and we find a 3$\sigma$ limiting column density limit of log $N_{\mathrm{NaI}}<9.45$. We see absorption in \ion{Ca}{2} across 6 pixels at 123.7 \kms\ and measure an AOD column density of log $N_{\mathrm{CaII}}$ = 10.68$\pm$0.01 for \ion{Ca}{2} 3934. We also performed a simultaneous Voigt profile fit to \ion{Ca}{2} $\lambda$3934, 3969 and find log $N$ = 10.63$\pm$0.10. The resulting $b$-value for this small component has a high error. However, since the log $N$ value from Voigt profile fitting agrees with the AOD measurement within the margin of error, we adopt its value.

\subsection{High ion absorption}

\begin{figure}[!ht]
    \centering
    \includegraphics[scale=0.52]{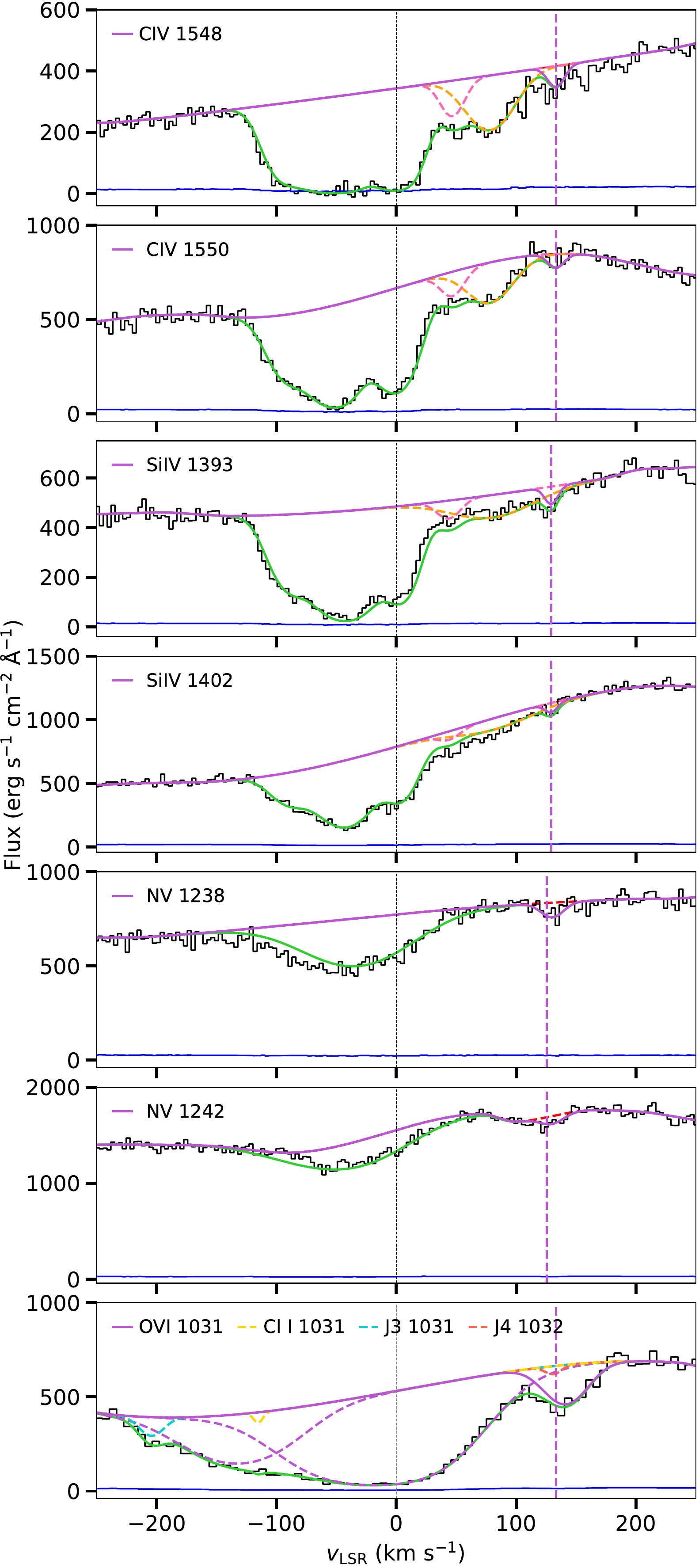}
    \caption{Velocity profiles of the high-ion absorption lines detected toward HD 156359. The flux is shown in black and the 1$\sigma$ error in the flux is in blue. The dashed black vertical line marks 0 km s$^{-1}$.  
    The solid green curve is the overall Voigt profile fit to the absorption. The solid magenta curve is the Voigt profile fit to the HVC absorption near $+125$ \kms\ and the vertical dashed magenta line marks the velocity centroid of the fitted component. The pink and yellow curves are fits to components at intermediate velocities.}
    \label{fig:vphigh}
\end{figure}

We detected \ion{C}{4} $\lambda\lambda$1548, 1550, \ion{Si}{4} $\lambda\lambda$1393, 1402, and \ion{N}{5} $\lambda\lambda$1238, 1242 absorption near $+$130 \kms\ in our STIS E140M spectrum.
The \ion{C}{4} and \ion{Si}{4} profiles are complex, showing multiple distinct low and intermediate-velocity components, in addition to the HVC. 
The high-ion absorption features lie on well-defined continua due to their corresponding broad and smooth stellar P Cygni wind profiles, which gradually elevates their stellar continua in the region from $-100$ to $+200$ \kms\ for this star, as shown in Figure \ref{fig:vphigh}. The high-ion continuum fitting was also guided by fits to the GHRS data previously published by 
\citet{sembach1991,sembach1995}. We performed Voigt-profile fitting to all high ions with the initial positions of the components and $b$-values determined from the \ion{C}{4} 1548 {\AA} line. The position and $b$-values were allowed to vary freely and the results for the fits are given in Table \ref{tab:colden}. We note the detection of an IVC centered near $+$78 \kms\ in both \ion{C}{4} and \ion{Si}{4} with log $N$ = $13.57\pm0.08$ and $13.09\pm0.10$, respectively.

We observe high-velocity absorption in \ion{O}{6} 1031 and 1037 in the LiF1 channel of the FUSE spectrum at \vlsr\ = 141 \kms. However, we only include $\lambda$1031 in our analysis, because $\lambda$1037 lies on the steep blueward side of the \ion{O}{6} P Cygni profile and suffers from significant blending with \ion{C}{2}$^{*}$ 1037 and H$_2$ $J$1 1038 {\AA}. 

The wavelength region around \ion{O}{6} 1031 {\AA} contains several absorption lines which can serve as contaminants, most notably HD 6--0 $R$(0) $\lambda$1031.915, \ion{Cl}{1} $\lambda$1031.507, and several lines of molecular hydrogen including H$_2$ 6--0 $P$(3) $\lambda$1031.192 and H$_2$ 6--0 $R$(4) $\lambda$1032.351. 
To gauge the effect of contamination by HD 6--0 $R$(0) $\lambda$1031.915, we examined other HD lines of similar oscillator strength that are isolated from interstellar absorption, such as 5--0 $R$(0) $\lambda$1042.850, 7--0 $R$(0) $\lambda$1021.460, and 8--0 $R$(0) $\lambda$1011.461. We detect no HD molecular absorption in these lines and conclude that no subtraction of the HD 6--0 $R$(0) line at $\lambda$1031.915 from the \ion{O}{6} profile is necessary. 
For \ion{Cl}{1}, a small amount of absorption in \ion{Cl}{1} $\lambda$1347 is present near 0 \kms\ in the STIS spectrum and we simultaneously fit \ion{Cl}{1} $\lambda$1031 to derive its contribution to the \ion{O}{6} profile.
For molecular hydrogen, H$_2$ 6--0 $R$(4) $\lambda$1032.351 is the most relevant potential contaminant as it overlaps with the high-velocity component in \ion{O}{6} $\lambda$ 1031. 
We modeled other isolated H$_2$ $J4$ lines in the FUSE spectrum, e.g. 5--0 $R$(4) $\lambda$1044.543 and 4--0 $R$(4) $\lambda$1057.381 (see description in Appendix \ref{sec:h2-decontaminate}). Through simultaneous fitting of those lines with H$_2$ 6--0 $R$(4) $\lambda$1032 , we account for the small amount of H$_2$ present in the \ion{O}{6} high-velocity absorption. A similar procedure was followed to account for contamination from H$_2$ 6--0 $P$(3) $\lambda$1031.192 near \vlsr\ = $-$200 \kms\ using the isolated H$_2$ $J3$ lines 6--0 $R$(3) $\lambda$1028.986 and 5--0 $P$(3) $\lambda$1043.503. The resulting Voigt profile fit and \ion{O}{6} column density are shown in Figure \ref{fig:vphigh} and Table \ref{tab:colden}.

\section{Results: The Cool Low-ion Gas}\label{sec:res_low}

The neutral and low ionization species including 
\ion{H}{1}, \oi, \ion{C}{2}, \ion{N}{2}, \ion{Si}{2}, \ion{Ca}{2}, \ion{Fe}{2}, \ion{Si}{3}, and \ion{Fe}{3} trace the cool photoionized phase of the HVC detected at \vlsr\ = +125 \kms\ toward \hd. In this section we use the ratios of metal column densities to \hi\ column densities to derive the ion abundances for each observed species in the HVC, as shown in Table \ref{tab:abund}. The elemental abundances are then derived from the ion abundances using ionization corrections derived from custom photoionization modeling. Finally a comparison of the relative abundances of different elements is used to derive the dust depletion pattern in the HVC. 

\begin{deluxetable}{lccc}[htb!]
\tablecaption{Measured Absorption Lines}
\label{tab:fdata}
\tablewidth{0pt}
\tablehead{
\colhead{Ion} & \colhead{$\lambda^a$} & \colhead{$f$} & \colhead{Dataset} \\
              & \colhead{(\AA)}        &               &                    }
\startdata
\hi          &  923.150 & 0.0022$^b$ & FUSE SiC2A\\
             &  926.226 & 0.0032$^b$ & FUSE SiC2A\\
             &  930.748 & 0.0048$^b$ & FUSE SiC2A\\
             &  937.804 & 0.0078$^b$ & FUSE SiC2A\\
\ion{C}{2}   & 1334.532 & 0.1290$^c$ & STIS E140M \\
\ion{C}{4}   & 1548.202 & 0.1900$^d$ & STIS E140M \\
             & 1550.774 & 0.0948$^d$ & STIS E140M \\
\ion{N}{2}   & 1083.990 & 0.1110$^c$ & FUSE SiC2B \\
\ion{N}{5}   & 1238.821 & 0.1560$^e$ & STIS E140M \\
             & 1242.804 & 0.0777$^e$ & STIS E140M \\
\oi\         & 1302.168 & 0.0480$^c$ & STIS E140M \\
\ion{O}{6}   & 1031.912 & 0.1330$^e$ & FUSE LiF1  \\
\ion{Na}{1}  & 5891.583 & 0.6500$^g$ & FEROS \\
\ion{Si}{2}  & 1190.420 & 0.2560$^f$ & STIS E140M \\
             & 1193.280 & 0.5440$^f$ & STIS E140M \\
             & 1304.370 & 0.0910$^f$ & STIS E140M \\
             & 1526.720 & 0.1440$^f$ & STIS E140M \\
\ion{Si}{3}  & 1206.510 & 1.6100$^g$ & STIS E140M \\
\ion{Si}{4}  & 1393.760 & 0.5130$^g$ & STIS E140M \\
             & 1402.770 & 0.2540$^g$ & STIS E140M \\
\ion{Ca}{2}  & 3934.770 & 0.6490$^h$ & FEROS \\
             & 3969.590 & 0.3210$^h$ & FEROS \\
\ion{Fe}{2}  & 1144.926 & 0.0830$^i$ & FUSE LiF2 \\
\ion{Fe}{3}  & 1122.524 & 0.0642$^j$ & FUSE LiF2  \\
\enddata
\tablecomments{
$^a$ All wavelengths are taken from \citet{morton2003}. 
$^b$ \citet{palchikov1998},
$^c$ \citet{froese2004},
$^d$ \citet{yan1998},
$^e$ \citet{peach1988},
$^f$ \citet{bautista2009},
$^g$ \citet{froese2006},
$^h$ \citet{safronova2011}
$^i$ \citet{donnelly2001},
$^j$ \citet{deb2009}}
\end{deluxetable}

\begin{deluxetable}{lccc}[ht!]
\tablecaption{HD\,156359 High-velocity absorption}
\label{tab:colden}
\tablewidth{0pt}
\tablehead{
\colhead{Ion} & \colhead{$v_\mathrm{LSR}$}  & \colhead{$b$}           & \colhead{log $N$}  \\
              & \colhead{(km s$^{-1}$)}     & \colhead{(km s$^{-1}$)} & \colhead{($N$ in cm$^{-2}$)}
}
\startdata
\ion{H}{1}   & 125.6 $\pm$ 0.9 & 18.9 $\pm$ 1.2 & 15.54 $\pm$ 0.05 \\
\ion{O}{1}   & 125.0 $\pm$ 4.8 & ...            & 12.43 $\pm$ 0.11$^a$ \\
\ion{Na}{1}  & ...             & ...            & $<$ 9.45$^b$ \\        
\ion{C}{2}   & 120.9 $\pm$ 0.6 &  9.8 $\pm$ 1.4 & 13.83 $\pm$ 0.10 \\
\ion{N}{2}   & ...             & ...            & $<$ 13.98$^b$ \\
\ion{Si}{2}  & 118.5 $\pm$ 1.0 &  7.0 $\pm$ 1.6 & 12.94 $\pm$ 0.07 \\ 
\ion{Ca}{2}  & 123.7 $\pm$ 1.0 &  2.3 $\pm$ 2.7 & 10.63 $\pm$ 0.10 \\ 
\ion{Fe}{2}  & ...             & ...            & $<$ 12.99$^b$    \\ 
\ion{Si}{3}  & 117.6 $\pm$ 1.4 &  6.9 $\pm$ 2.8 & 12.32 $\pm$ 0.16 \\
\ion{Fe}{3}  & 129.3 $\pm$ 2.3 & 11.0 $\pm$ 3.6 & 12.82 $\pm$ 0.14 \\
\ion{C}{4}   & 133.3 $\pm$ 2.3 &  8.6 $\pm$ 3.6 & 12.60 $\pm$ 0.13 \\
\ion{Si}{4}  & 129.5 $\pm$ 1.5 &  5.7 $\pm$ 2.9 & 11.93 $\pm$ 0.15 \\
\ion{N}{5}   & 130.2 $\pm$ 2.7 & 12.2 $\pm$ 4.0 & 12.64 $\pm$ 0.11 \\
\ion{O}{6}   & 141.3 $\pm$ 1.2 & 24.0 $\pm$ 1.8 & 13.65 $\pm$ 0.03 \\
\enddata
\tablecomments{
$^a$ Except where noted, all measurements presented in this table are derived from Voigt-profile fitting, with the exception of \oi, for which we derive an apparent column density from the profile in the range $+110 \lesssim v \lesssim +125$ km s$^{-1}$, see Figure \ref{fig:nav}. The error listed for log\,$N$(\oi) is the combination of the statistical error and a continuum placement error in quadrature, and is $\pm$ 0.08 $\pm$ 0.07 separately. \\
$^b$ This measurement is a non-detection and the 3$\sigma$ limiting column density is given.
}
\end{deluxetable}

\subsection{Photoionization Modeling}\label{subsec:cloudy}

To model the ionization breakdown in the HVC and characterize its physical conditions, we ran a multi-dimensional grid of \textsc{Cloudy} (v.\,17.02, \citealt{ferland2017}) photoionization models for the low- and intermediate-ions, assuming they arise in the same gas phase as the \ion{H}{1} and are photoionized by the same incident radiation field. Our model assumes a plane-parallel slab of uniform density exposed to the magnitude of the escaping UV ionizing flux of the Milky Way \citep{fox2005,barger2013,fox2014} at the location of \hd\ and includes the extragalactic UV background radiation field from \citet{khaire2019}.

\begin{figure*}[!ht]
    \centering
    \includegraphics[scale=0.39]{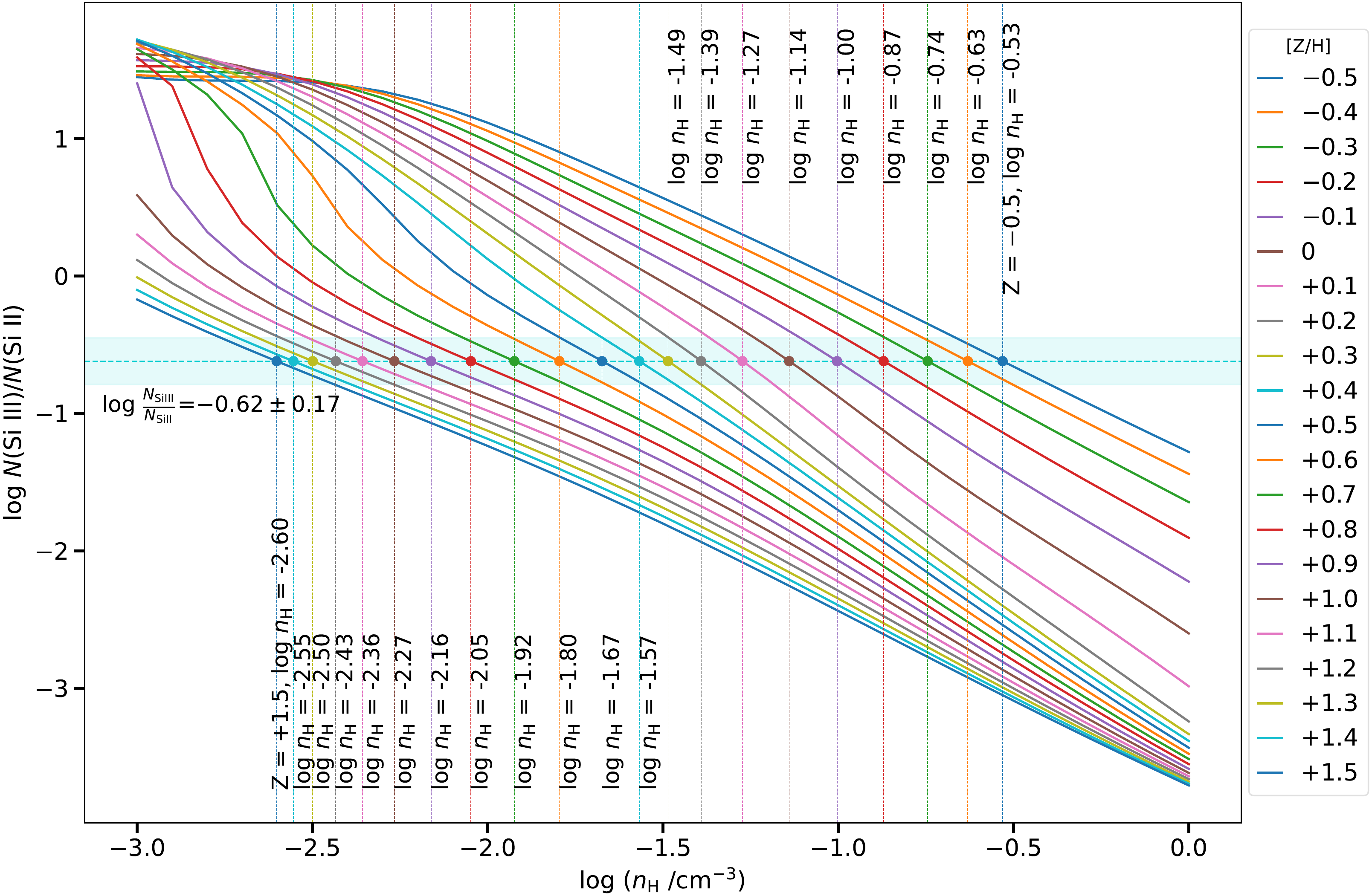}
    \includegraphics[scale=0.39]{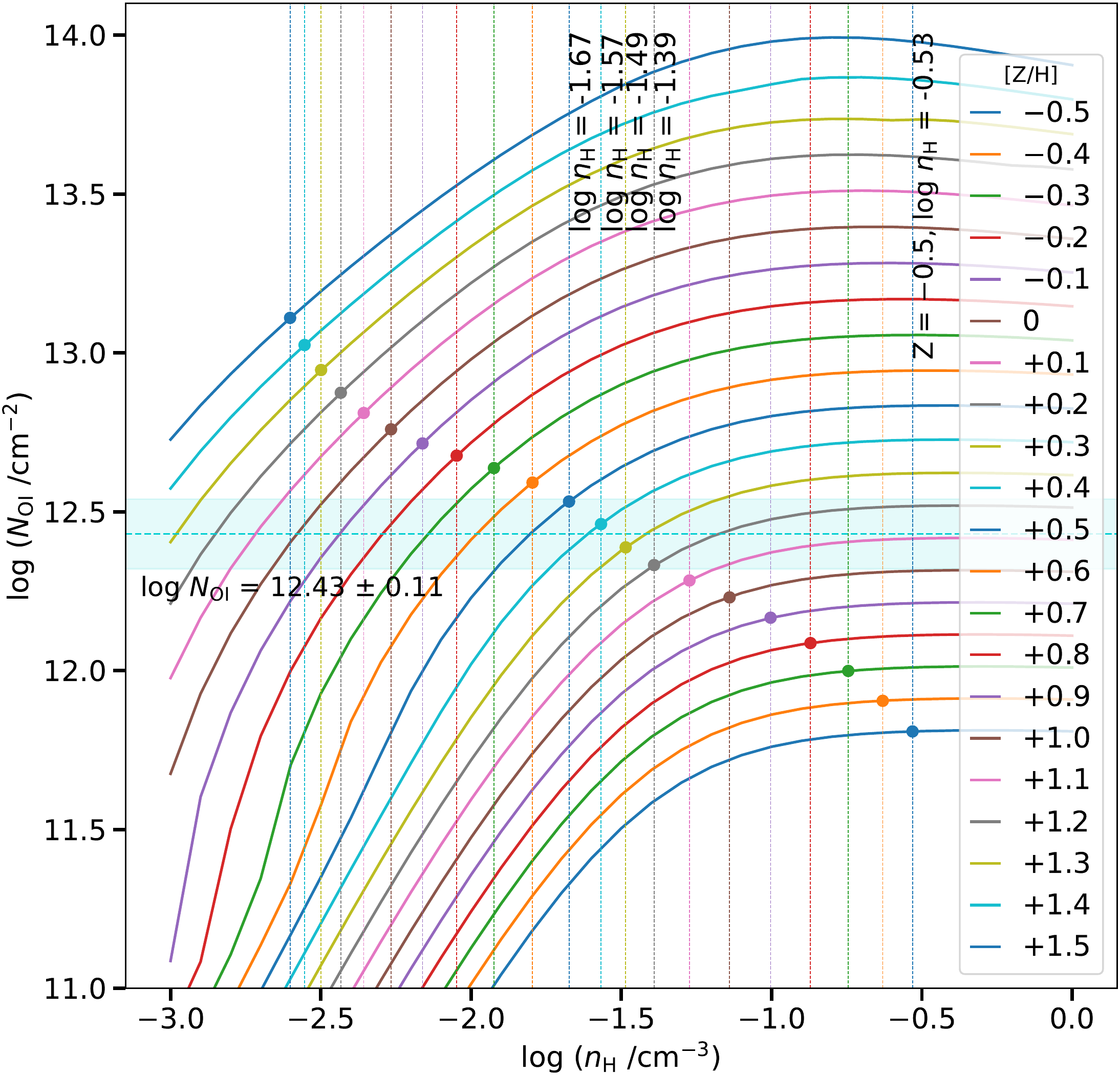}
    \includegraphics[scale=0.39]{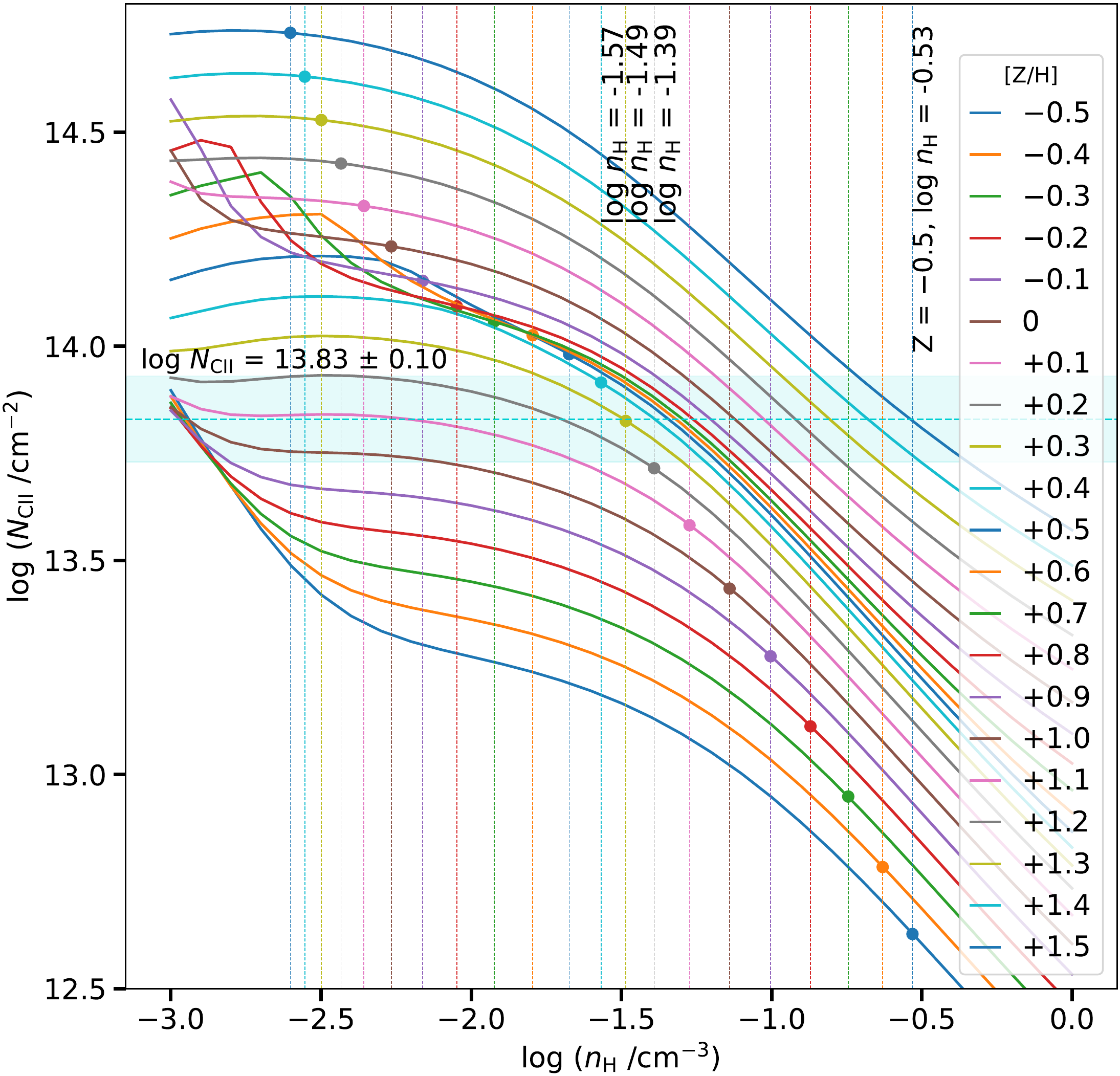}
    \caption{\cloudy\ photoionization models of the HVC toward \hd\ that explore a parameter space in metallicity and number density, $-0.50 \leq$ [Z/H] $\leq +1.50$ and $-3 \leq$ log ($n_\mathrm{H}$/cm$^{-3}$) $\leq 0$, respectively. Top panel: Model results for log $N_\mathrm{SiIII}$/$N_\mathrm{SiII}$ for the set metallicities are shown as solid colored curves. The dashed horizontal turquoise line is the observed ratio of $-$0.62$\pm$0.17. The best-fit log $n_\mathrm{H}$ for each metallicity model is determined by the intersection of the model and observed ratio and is marked by vertical colored lines and circle markers. Bottom left panel: The model curves for log $N_\mathrm{OI}$ for each metallicity are plotted against \logh, where the values derived from 
    log~$N_\mathrm{SiIII}$/$N_\mathrm{SiII}$ are marked with vertical lines and markers. The horizontal turquoise line and bar show the observed 
    log~$N_\mathrm{OI}$ = 12.43 $\pm$ 0.11. The model data corresponding to the metallicity region from $+0.18 \leq$[Z/H] $\leq +0.51$ and $-1.69\leq$ \logh\ $\leq-1.37$ overlap with the observed data. The model best-fit values are [Z/H] = $+$0.36$^{+0.15}_{-0.18}$ and \logh\ = $-1.53\pm0.16$. Bottom right panel: Same as for the bottom left panel, but for \ion{C}{2}. The horizontal turquoise line and bar show observed log $N_\mathrm{CII}$ = 13.83 $\pm$ 0.10. The model data corresponding to the metallicity region from $+0.21 \leq$ [Z/H] $\leq +0.42$ and $-1.59\leq$ \logh\ $\leq-1.40$ overlap with the observed data. The model best fit values are [Z/H]= $+$0.30$^{+0.12}_{-0.09}$ and \logh\ = $-1.49^{+0.09}_{-0.10}$.}
    \label{fig:cloudy1}
\end{figure*}

We ran our models for a grid of metallicity values with [Z/H] varying from $-0.5$ to $+1.5$ in steps of 0.1 dex, each over a range of hydrogen number densities log ($n_\mathrm{H}$/cm$^{-3}$) from $-3$ to 0, in order to explore possible metallicities and densities. We determined the best-fit log $n_\mathrm{H}$ for each metallicity model by matching the observed column density ratio of \ion{Si}{3}/\ion{Si}{2} to the model value (see the top panel of Figure \ref{fig:cloudy1}). The \ion{Si}{3}/\ion{Si}{2} ratio was chosen because both ions have unsaturated detections in the HVC, and using a ratio of adjacent ions from the same element (Si) minimizes metallicity or depletion effects from different metal ions. Using the pairs of metallicity and \logh\ determined from the \ion{Si}{3}/\ion{Si}{2} ratio, we narrow the range of metallicities by identifying which models are consistent with the observed log $N_\mathrm{OI}$ = 12.43 $\pm$ 0.11, as illustrated in the bottom left panel of Figure \ref{fig:cloudy1}. We then run an even finer grid of metallicity values in increments of 0.02 dex over the narrowed metallicity and number density region to determine the range of metallicities and densities allowed by the data. 

We find that metallicities in the range $0.18 \leq \mathrm{[O/H]} \leq 0.51$ and densities from $-1.69 \leq$ log ($n_\mathrm{H}$/cm$^{-3}$) $\leq -1.37$ are consistent with the data, giving a model best-fit of [O/H] = 0.36$^{+0.15}_{-0.18}$ at log $n_\mathrm{H}$ = $-$1.53$\pm$-0.16. We repeated this procedure for \ion{C}{2} (see bottom-right panel of Figure \ref{fig:cloudy1}), which is expected to be weakly depleted \citep{jenkins2009}, and find that the range 0.21 $\leq$ [C/H] $\leq$ 0.42 overlaps with the observed data, giving a model best-fit of [C/H] = 0.30$^{+0.12}_{-0.09}$ at \logh\ = $-$1.49$^{+0.09}_{-0.10}$. 
The agreement between the oxygen-based metallicity and the carbon-based metallicity lends support to our methodology and to the robustness of the super-solar metallicity we infer.

The metallicity for this cloud ranks among the highest UV-based metallicities of any HVC observed thus far, along with the HVC at $-$125 \kms\ toward M5-ZNG ($l$ = 3.$\degree$9, $b$ = $+$47.$\degree$7 at $z$ = $+$5.3 kpc) with [O/H] = $+$0.22 $\pm$ 0.10 \citep{zech2008}. M5-ZNG was also observed with FUSE and STIS E140M spectra. Their observed metallicity is not corrected for ionization effects, but is likely higher than reported, given the measured low \loghone\ = 16.50 $\pm$ 0.06 in the cloud and that oxygen shows a positive ionization correction  when log\,$N$(\ion{H}{1}) $<$ 18.5 \citep{bordoloi2017}. The metallicity of the \hd\ HVC is also on the high end of the range of $<$20\% to 3 times solar reported by \citet{ashley2022} in their survey of metallicities of Fermi Bubble HVCs. We note that a super-solar abundance in the inner Galaxy may not be unexpected, given the oxygen abundance gradients reported from emission line measurements in \ion{H}{2} regions (see \citealt{wenger2019, arellano2021}).

\begin{deluxetable}{lcccc}[!ht]
\tablecaption{HVC elemental abundances and depletions}
\label{tab:abund}
\tablewidth{0pt}
\tablehead{
\colhead{Ion} & \colhead{[X$^i$/\ion{H}{1}]$^a$} & \colhead{IC(X$^i$)$^b$} &  \colhead{[X/H]$^c$} & \colhead{$\delta_\mathrm{O}$(X)$^d$} 
}
\startdata
\oi\         & 0.20$\pm$0.10 & $+$0.16$^{+0.04}_{-0.07}$  & 0.36$^{+0.11}_{-0.12}$    & 0 \\
\ion{C}{2}   & 1.86$\pm$0.12 & $-$1.59$\pm$0.04           & 0.27$\pm$0.13             & $-$0.09$\pm$0.17 \\
\ion{N}{2}   & $<$ 2.60      & $<-$1.67                   & $<$ 0.93                  & $<$ 0.57 \\  
\ion{Si}{2}  & 1.89$\pm$0.08 & $-$1.86$\pm$0.03           & 0.03$\pm$0.09             & $-$0.33$\pm$0.14 \\
\ion{Fe}{2}  & $<$ 1.95      & $-$1.17$^{+0.20}_{-0.12}$  & $<$ 0.79                  & $<$ 0.43 \\
\ion{Fe}{3}  & 1.78$\pm$0.15 & $-$1.73$\pm$0.08           & 0.06$\pm$0.17             & $-$0.30$\pm$0.20 \\
\ion{Ca}{2}  & 0.75$\pm$0.11 & $-$0.95$\pm$0.01           & $-$0.20$\pm$0.11          & $-$0.56$\pm$0.16 \\
\enddata
\tablecomments{\\
$^a$ Ion abundance [X$^i$/\ion{H}{1}] = log (X$^i$/\ion{H}{1})$_\mathrm{HVC}$ $-$ log (X/H)$_\odot$ where X$^i$ is the observed ion of element X. These are not corrected for ionization effects. The error includes both systematic and continuum placement uncertainties. \\
$^b$ Ionization correction IC(X$^i$) = [X/H]$_\mathrm{model}$ $-$ [X$^i$/\ion{H}{1}], derived from the difference between the model elemental abundance and the observed ion abundance. \\ 
$^c$ Ionization-corrected elemental abundance [X/H] = [X$^i$/\ion{H}{1}] + IC(X$^i$). Also referred to as gas-phase abundance.\\
$^d$ $\delta_\mathrm{O}$(X) $\equiv$ [X/O] = [X/H] $-$ [O/H] is the depletion of element X relative to oxygen.
}
\end{deluxetable}

\begin{figure}[!ht]
    \centering
    \includegraphics[scale=0.37]{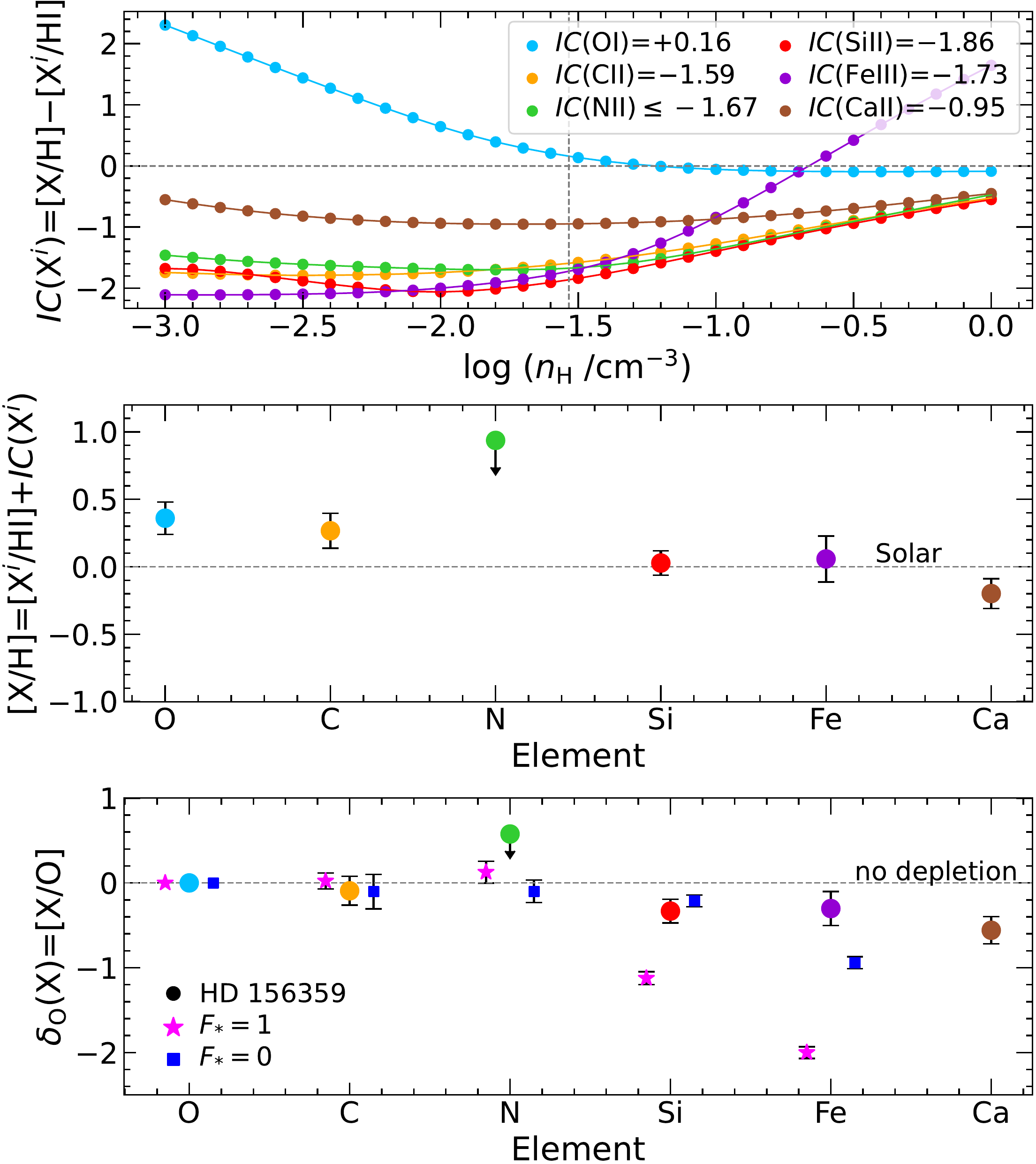}
    \caption{Top panel: ionization corrections for the low ions \oi, \ion{C}{2}, \ion{N}{2}, \ion{Si}{2}, \ion{Fe}{3}, and \ion{Ca}{2} (colored curves) plotted against \logh\ from our \cloudy\ model to the \hd\ HVC. The model uses our best-fit [Z/H]=$+$0.36. The vertical line marks the best-fit log\,($n_{\rm H}$/cm$^{-3}$)=$-$1.53, determined from the \cloudy\ models for \oi\ (see Figure \ref{fig:cloudy1}). The ionization correction calculated at $-$1.53 is added to the observed ion abundance to determine the elemental abundance. Middle panel: comparison of the ionization-corrected abundances of the low ions detected in the HVC, where the solar abundance is plotted as a horizontal line. Bottom panel: Comparison of the dust depletion levels $\delta$(X) = [X/O] in the HVC with the depletion pattern for [X/O] measured for sight lines with the lowest and highest collective depletions ($F_{*}$=0 and $F_{*}$=1, respectively) from \citet{jenkins2009}. A slight offset is applied in the $x$-direction of each element for distinction.}
    \label{fig:ic_met_dep}
\end{figure}

\begin{figure}[!ht]
    \centering
    \includegraphics[scale=0.45]{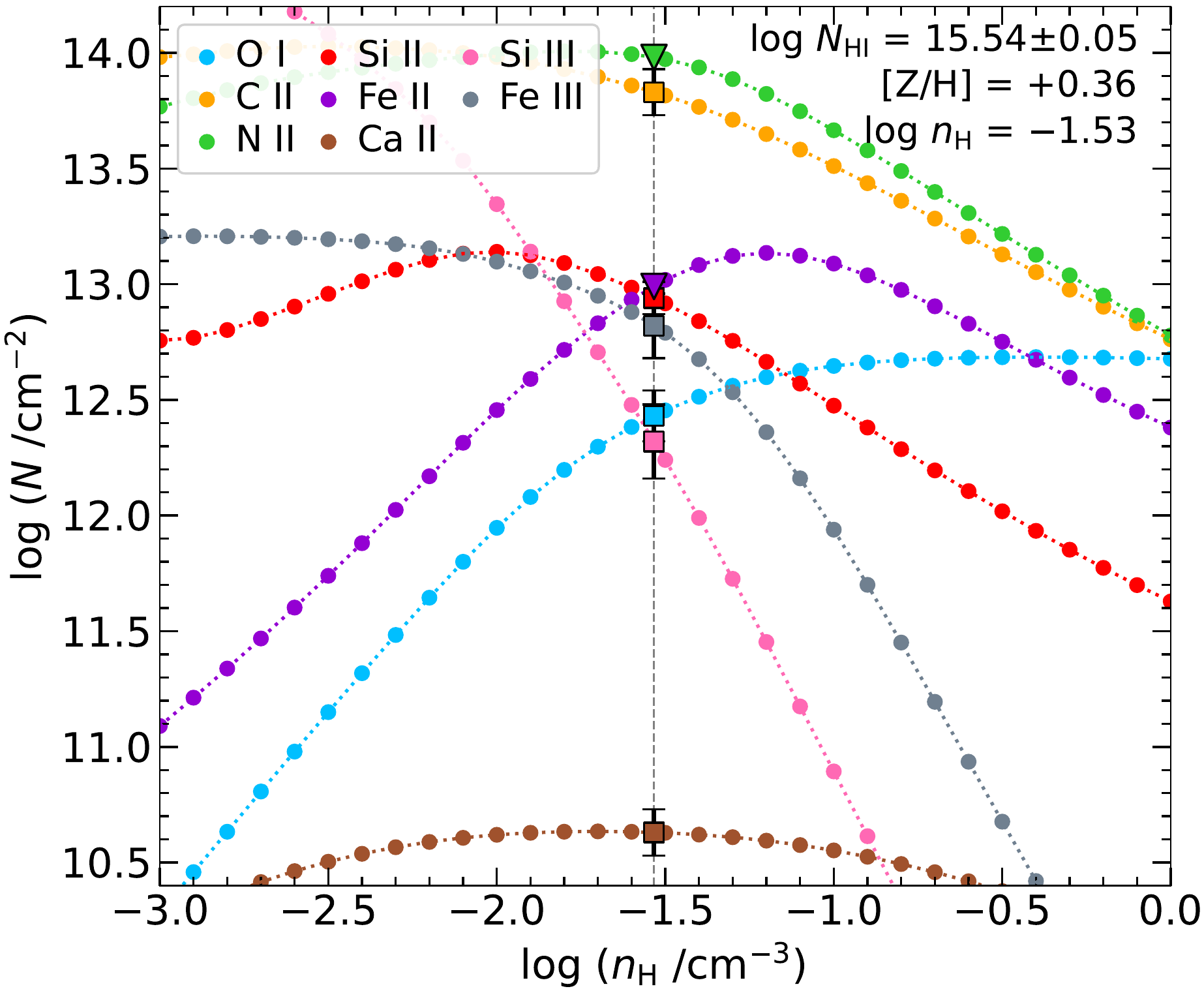} 
    \caption{Our final \textsc{Cloudy} photoionization model of the low- and intermediate ions detected
    in the HVC toward HD 156359 near $+125$~\kms. The model uses the best-fit [Z/H] = $+$0.36 determined from our \oi\ analysis. 
    These model predictions for each ion (colored dotted curves) have been scaled by the dust depletions required to match the observed values, where the observed column densities are indicated by circle markers at the best-fit log ($n_\mathrm{H}$/cm$^{-3}$)=$-$1.53 (black vertical line). 
    }
    \label{fig:cloudy2}
\end{figure}

\subsection{Ionization corrections}\label{sec:ic}

In ISM abundance work, [\oi/\hi] is often considered a robust indicator of elemental abundance [O/H], since (1) charge-exchange reactions closely couple the ionization state of hydrogen and oxygen together \citet{fs71}, 
and (2) oxygen is only lightly depleted onto dust grains \citep{jenkins2009}. 
However, the assumption that [\oi/\hi]=[O/H] breaks down when $N$(\hi) is so low that the gas is optically thin or when the ionizing photon flux is extremely high \citep{viegas1995}.
In the HVC at +125~\kms\ toward \hd, log\,$N$(\hi) is only 15.54$\pm$0.05, so an ionization correction must be made to account for the column densities of all observed neutral and ion stages.

Our \textsc{Cloudy} models directly provide the ionization corrections for all our observed ion stages. We define the ionization correction as the difference between the model (true) elemental abundance and the observed ion abundance, i.e.,
\begin{equation}
    IC(X^i)=[\text{X/H}]-[\text{X}^i/\textsc{H i}].
\end{equation}

The resulting ion abundances, ionization corrections, and ionization-corrected elemental abundances are given in Table \ref{tab:abund}. The relationships between the ionization corrections and the hydrogen number density for [Z/H] = 0.36 are shown in the top panel of Figure \ref{fig:ic_met_dep} and the ionization-corrected abundances are shown in the middle panel of Figure \ref{fig:ic_met_dep}. 

The positive ionization correction of $IC$(\oi) = $0.16^{+0.04}_{-0.07}$ from the photoionization modeling results in a high gas-phase oxygen abundance of $+0.36$. We calculate gas-phase abundances of [C/H] = 0.27$\pm$0.13, [N/H] = $\leq$ 0.93, [Si/H] = 0.03$\pm$0.09, [Fe/H] = 0.06$\pm$0.17, and [Ca/H] = $-$0.20$\pm$0.11 (see Table~\ref{tab:abund}).

\subsection{Dust depletion effects}\label{sec:depl}

Following convention, we define the depletion $\delta_\mathrm{O}$(X) of each refractory element X as the ionization-corrected abundances of that element compared to the ionization-corrected oxygen abundance, i.e.,
\begin{equation}
    \delta_\mathrm{O}(\text{X}) \equiv [\text{X/O}] = [\text{X/H}] - [\text{O/H}],
\end{equation}

where oxygen represents an undepleted (or lightly depleted) volatile element.
In this formalism, a negative value of $\delta_\mathrm{O}$(X) means that element X is depleted relative to oxygen.
This method assumes that the total (gas+dust) abundances are solar, which is often assumed to apply to the Galactic ISM \citep{savage1996}, though local ISM abundance variations cannot be ruled out in some sight lines \citep{decia2021}.
We show $\delta_\mathrm{O}$(X) for the low ions in the bottom panel of Figure \ref{fig:ic_met_dep} 
compared to $F_{*}$, the line-of-sight depletion strength factor, for [X/O] determined for low-depletion ($F_{*}$=0) and high-depletion ($F_{*}$=1) gas from the comprehensive ISM gas-phase element depletions study of \citet{jenkins2009}. 

For the HVC toward \hd, we find a low value of $\delta_\mathrm{O}$(C)=$-$0.09$\pm$0.17, i.e. carbon shows no significant depletion. This is consistent with the low values of [C/O] measured in Galactic ISM gas \citep{jenkins2009}. 
For the other low ions, we find a 3$\sigma$ upper limit for the nitrogen depletion $\delta_\mathrm{O}$(N)$\leq+0.57$ and low depletions for silicon, iron, and calcium of  
$\delta_\mathrm{O}$(Si)=$-$0.33$\pm$0.14$, \delta_\mathrm{O}$(Fe)=$-$0.30$\pm$0.20, and
$\delta_\mathrm{O}$(Ca)=$-$0.56$\pm$0.16, respectively.
To summarize the final results of the CLOUDY modeling after the dust depletion levels have been derived, Figure \ref{fig:cloudy2} shows the model curves of the detected ions shifted by 
their respective depletions at [X/H] = $+$0.36 and \logh = $-$1.53. 
The ability of this model to match the observed column densities shows that all low-ion measurements can be explained by photoionization once dust depletion effects are accounted for.

A low level of dust depletion for the HVC is consistent with the well-known Routly-Spitzer effect (RS effect), in which the amount of dust observed in high-velocity clouds decreases significantly at higher LSR velocity (see \citealt{routly1952, siluk1974, vallerga1993, smoker2011, benbekhti2012, murga2015}). The RS effect is typically traced by the $N$(\ion{Na}{1})/$N$(\ion{Ca}{2}) ratio, which has been found to significantly decrease with increasing LSR velocity. Although the RS effect was historically interpreted as a dust depletion effect, it may also be related to ionization effects, and likely differs depending on the environment of the cloud. 
We measure an 3$\sigma$ upper limit of log $N$(\ion{Na}{1})/$N$(\ion{Ca}{2}) $\leq$ $-$1.2 in the $+124$ \kms\ HVC toward \hd\ from the optical FEROS data, which is on the low end of the observed ratios for HVCs \citep{smoker2011, benbekhti2012, murga2015}, and is also  
lower than is typically observed at low velocities. In summary, the dust depletion pattern we derive in the HVC is consistent with other HVCs and with the Routly-Spitzer effect, even though the cloud has high metallicity.

\section{Results: the Highly Ionized Gas}\label{sec:res_high}

\begin{figure}[!t]
    \centering
    \includegraphics[scale=0.38]{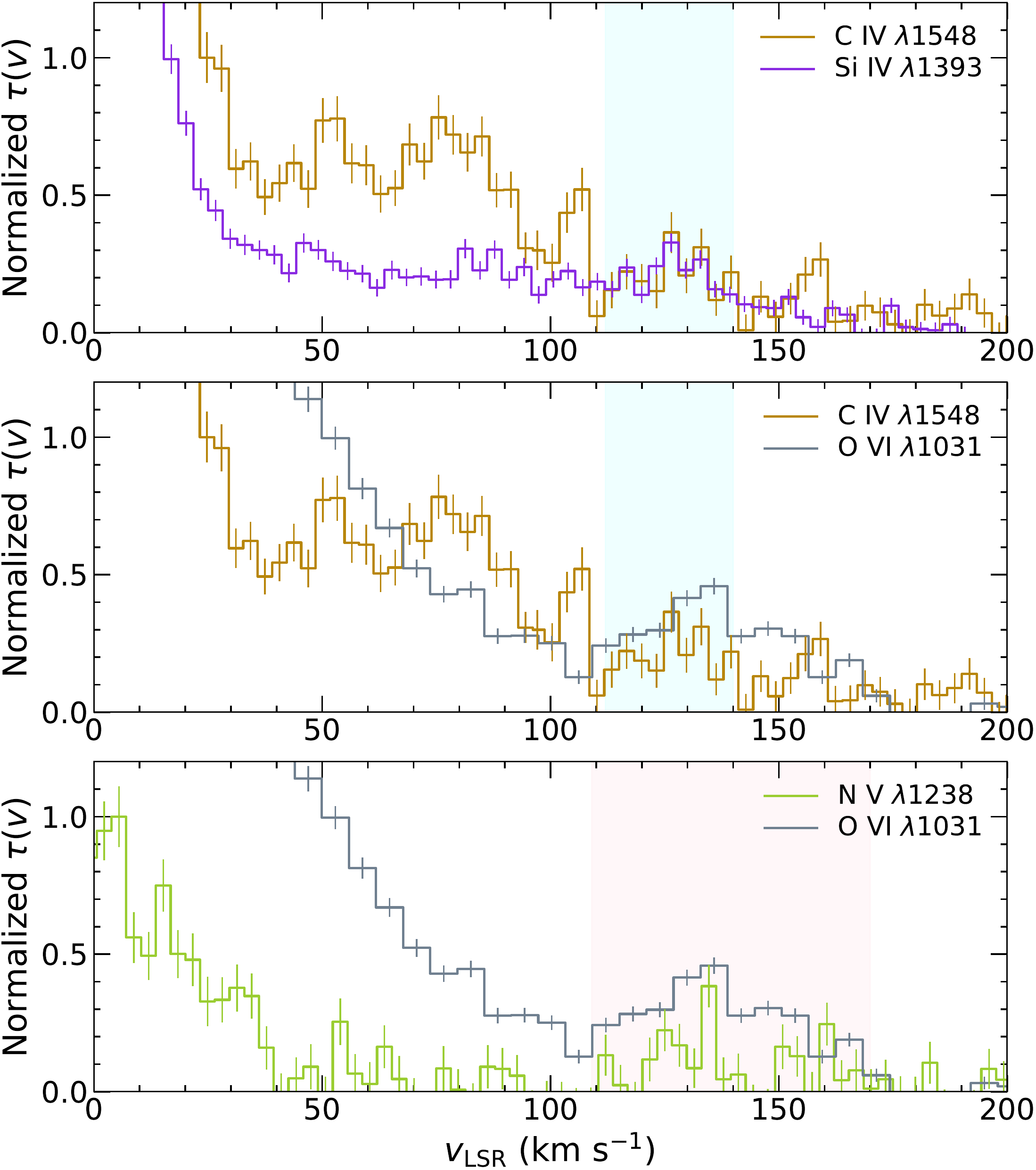}
    \caption{Optical depth profiles of the high ions, where each profile is arbitrarily normalized to its peak value for the velocity region shown in order to facilitate comparison. Top panel: \ion{C}{4} $\lambda$1548 and \ion{Si}{4} $\lambda$1393 display similar line shapes in the velocity region of the HVC from $112 \lesssim v \lesssim 140$ \kms, indicated with cyan shading. Middle panel: \ion{C}{4} $\lambda$1548 and \ion{O}{6} $\lambda$1031 are compared, where the center of the HVC absorption for \ion{O}{6} is shifted to a higher velocity at $v\approx141$ \kms. Bottom panel: \ion{N}{5} $\lambda$1238 and \ion{O}{6} $\lambda$1031 are compared, where the range of velocities attribute to absorption by \ion{O}{6} are indicated in light pink.}
    \label{fig:tau}
\end{figure}

The HVC toward \hd\ shows high-ion absorption in the resonance doublets of 
\ion{Si}{4}, \ion{C}{4}, \ion{N}{5}, and \ion{O}{6} at a velocity range of 130--140~\kms\
(Figure \ref{fig:vphigh}; Table \ref{tab:colden}),
slightly higher than the velocity range of the low ions, which are centered near 125 \kms.
To compare the high-ion absorption between different ions, we show in Figure \ref{fig:tau} a comparison of 
the normalized optical depth profiles of each high ion. 
These plots provide a way to inter-compare the profiles of different ions in a linear manner (e.g., \citealt{fox2003}).
The HVC line profiles of \ion{C}{4} and \ion{Si}{4} are well aligned in velocity, but are offset from the centroid of \ion{O}{6} absorption by $\approx$8 \kms.
This suggests that the \ion{O}{6} exists in a separate (likely hotter) gas phase. 
The STIS profile of \ion{N}{5} is too noisy to draw any conclusions about line centroid. However, the GHRS \ion{N}{5} profiles \citep{sembach1995} are less noisy and have a similar profile shape to the \ion{O}{6} seen in the FUSE data, supporting the placement of \ion{O}{6} and \ion{N}{5} in the same phase. 

\subsection{Collisional Ionization Modeling}\label{subsec:orly}

\begin{figure}[!htb]
    \centering
    \includegraphics[scale=0.44]{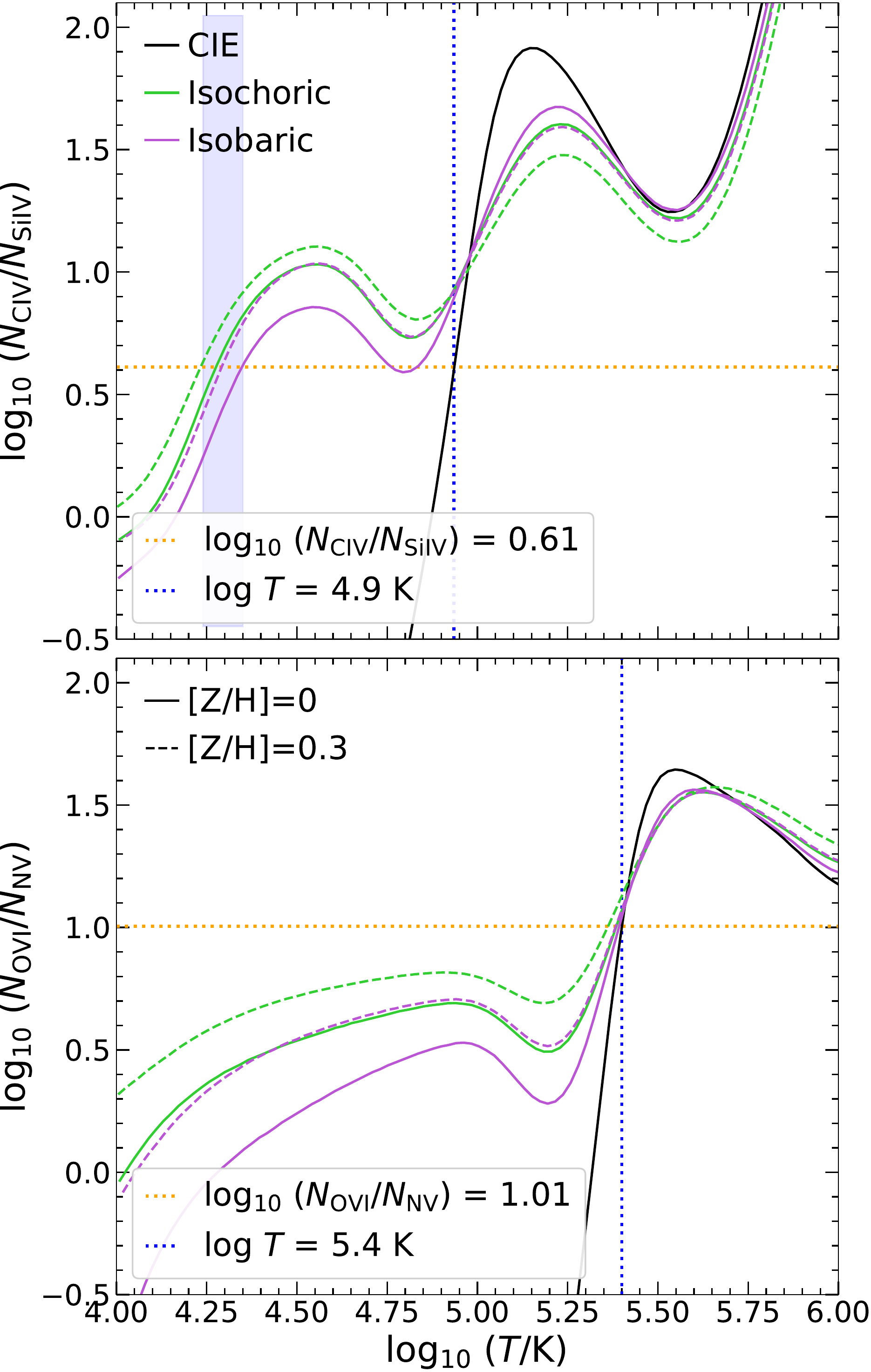}
    \caption{Comparison of observed high-ion column-density ratios with predictions from the 
    collisional ionization models of \citet{gnat2007}. 
    The observed ratios in the \hd\ HVC are shown as dashed orange horizontal lines.
    The top and bottom panels show the \ion{C}{4}/\ion{Si}{4} and \ion{O}{6}/\ion{N}{5} ratios, respectively. In both panels, the solid black curve is the collisional ionization equilibrium model ion fraction. The solid green and violet curves are the time-dependent isochoric and isobaric collisional ionization models for a solar metallicity, respectively. Their dashed counterparts are the same but for a metallicity twice the solar value. The vertical dashed blue line is the temperature of the gas determined by the models. The vertical blue band in the top panel represents a second time-dependent solution for a range of temperatures from log $T$ = 4.2--4.4 K.}
    \label{fig:orly}
\end{figure}

The observed high ions have column densities that are too high to be explained 
by the \textsc{Cloudy} photoionization models described in Section~\ref{subsec:cloudy}. Where the \ion{Si}{4} and \ion{C}{4} column densities differ by $\sim$0.7 and 1.0 dex respectively, the \ion{N}{5} and \ion{O}{6} differ by orders of magnitudes. These discrepancies imply that a separate ionization mechanism is required for the high ions. We do not account for energetic photons from the radiation field intrinsic to the Fermi Bubbles, however, we explore collisional ionization as a separate mechanism for gas with temperatures of $T \gtrsim$ 10$^{5}$ K. We consult the collisional ionization models from \citet{gnat2007} to determine the exact range of temperatures allowed by the high-ion column densities and their ratios. While we considered the more recent models from \citealt{gnat2017}, which include photoionization effects from the extragalactic UV background, we do not adopt them because HD 156359 lies in close proximity to the radiation field of the Galactic plane which is much stronger. We considered both the collisional ionization equilibrium and non-equilibrium regimes.

A single-temperature solution explaining all four high ions 
(\ion{Si}{4}, \ion{C}{4}, \ion{N}{5}, \ion{O}{6}) in the HVC is ruled out, 
as no such solution exists to the observed high-ion column densities. 
Instead, we find that a two-phase solution is needed, with one temperature explaining the $N_\mathrm{CIV}$/$N_\mathrm{SiIV}$ ratio and another explaining $N_\mathrm{OVI}$/$N_\mathrm{NV}$. This two-phase model is consistent with the kinematic information in the UV spectra, 
where \ion{Si}{4} and \ion{C}{4} show very similar line profiles but \ion{O}{6} is offset in velocity centroid.
The two-phase high-ion model is illustrated in Figure \ref{fig:orly}, which shows the observed high-ion ratios of $N_\mathrm{CIV}$/$N_\mathrm{SiIV}$ and $N_\mathrm{OVI}$/$N_\mathrm{NV}$ 
compared to model predictions from \citet{gnat2007} for solar ([Z/H]=0) and super-solar ([Z/H]=$+$0.3) metallicities. 

For the \ion{C}{4}/\ion{Si}{4} phase, we find two possible solutions for the temperature. First, the non-equilibrium isochoric (constant volume) and isobaric (constant pressure) models give a low-temperature solution at $T$ = $10^{4.2-4.4}$ K, where the lower end corresponds to the solar-metallicity isochoric model and the higher end with the super-solar isobaric model. Second, the collisional ionization equilibrium (CIE) model returns a higher temperature $T$=10$^{4.9}$ K. We are inclined to adopt the non-equilibrium (lower-temperature) solution, as plasma near $T$=10$^{5}$ K is at the peak of the cooling curve, where oxygen dominates the radiative cooling and the gas can cool faster than it recombines, reaching a non-equilibrium state. 
For the \ion{O}{6}/\ion{N}{5} phase, a single temperature solution for the observed 
log $N_\mathrm{OVI}$/$N_\mathrm{NV}$=1.01 
is found at $T$=10$^{5.4}$ K for all models (see bottom panel of Figure \ref{fig:orly}). 

In conclusion, we can successfully model the high-ion plasma in the HVC toward \hd\ as containing collisionally-ionized gas at two temperatures: a cooler phase seen in \ion{C}{4} and \ion{Si}{4} at $T$ = $10^{4.2-4.4}$ K, and a hotter phase seen in \ion{N}{5} and \ion{O}{6} at $T$=10$^{5.4}$ K. 

\section{Summary} \label{sec:sum}

Using archival FUSE, HST STIS, and ESO FEROS spectra, we have analyzed the chemical composition of the HVC near $+$125 \kms\ toward \hd, a massive star lying 9~kpc away toward the Galactic Center. The sight line passes less than 1$\degree$ from one of the densest cores of a complex of small HVCs dubbed ``Complex WE'' by \citet{wakker1991}, as shown in Figure \ref{fig:allsky}.
Furthermore, the sight line passes through a region of enhanced X-ray emission \citep[the southern eROSITA Bubble;][]{predehl2020}; this region indicates energetic feedback from the Galactic Center. Our main results are as follows.

\begin{enumerate}

\item We determined an \hi\ column density measurement of log\,$N$(\hi) = $15.54\pm0.05$ in the HVC using unsaturated Lyman series absorption lines.

\item We measured a super-solar \ion{O}{1} abundance of [\oi/\ion{H}{1}] = 0.20$\pm$0.11 in the HVC. After applying an ionization correction, we derive an oxygen abundance of [O/H] = 0.36$\pm$0.12, indicating the cloud is enriched to 2.3 times the solar level. This abundance is consistent with super solar O abundances of \ion{H}{2} regions measured in the inner Galaxy \citep{wenger2019, arellano2021}.

\item The ionization-corrected carbon abundance [C/H] = 0.27$\pm$0.13 is consistent with the oxygen abundance. 
This indicates that carbon shows low dust depletion relative to oxygen, consistent with the pattern seen by \citet{jenkins2009}, as shown in Figure \ref{fig:ic_met_dep}. 

\item A low level of depletion is inferred from the silicon, iron, and calcium lines, with [Si/O] = $-$0.33 $\pm$ 0.14, [Fe/O] = $-$0.30 $\pm$ 0.20, and [Ca/O] = $-$0.56 $\pm$ 0.16. The \ion{Na}{1}/\ion{Ca}{2} ratio in the HVC is $\leq-1.2$, which is lower than is typically observed at low velocities, consistent with the trend known as the Routly-Spitzer effect.

\item We detect high ion species \ion{C}{4}, \ion{Si}{4}, \ion{N}{5}, and \ion{O}{6} near $+$130 \kms, at slightly higher velocities than the velocity range of the lower ions. We find that an ionization mechanism separate from photoionization, such as collisional ionization, is required to explain the column densities of the high ions. We determine that a two-phase temperature solution best explains the observed  $N_\mathrm{CIV}$/$N_\mathrm{SiIV}$ and $N_\mathrm{OVI}$/$N_\mathrm{NV}$ ratios, with a cooler phase seen in \ion{C}{4} and \ion{Si}{4} at $T$ = $10^{4.2-4.4}$ K, and a hotter phase seen in \ion{N}{5} and \ion{O}{6} at $T$=10$^{5.4}$ K.

\end{enumerate}

The high metallicity, low depletion, complex high-ion absorption, and high positive velocity of the HVC toward \hd\ are all consistent with a wind origin, in which a swept-up clump of material is being carried out from the Galactic disk into the halo. 
As such, this HVC may represent a freshly entrained cool clump of gas caught in the act of being accelerated into the nuclear wind. 
While we cannot rule out a foreground origin for the HVC, in which the cloud exists at an anomalous velocity somewhere between the Sun and the Galactic Center, we can exclude a spiral-arm explanation for the HVC on kinematic grounds, because the cloud's $+125$ \kms\ velocity lies over 100 \kms\ away from the nearest spiral arm, the Sagittarius Arm at $\sim-10$ \kms. 
Under a biconical outflow model (\citealt{fox2015, bordoloi2017, diteodoro2018}), the cloud's radial velocity and its location only 2.3 kpc below the disk imply a short timescale of $\sim$5 Myr for the HVC to have reached its current position, which is much shorter than the timescales on which chemical mixing is expected to be significant \citep[10s to 100s of Myr;][]{gritton2014,heitsch2022}. 
Our observations therefore provide a snapshot into the chemical and physical conditions prevailing in this early stage of a nuclear outflow, before chemical mixing has diluted or enriched the gas from its initial state.


\begin{acknowledgments}
We gratefully acknowledge the invaluable contributions to early versions of this manuscript from the late Blair Savage, who passed away during the manuscript's preparation. This paper would not have been possible without Blair's foundational work on the \hd\ sight line, the chemical abundances in the ISM, apparent optical depth analysis, and the inner Galaxy. 
We gratefully acknowledge support from the NASA Astrophysics Data Analysis Program (ADAP) under grant 80NSSC20K0435, 3D Structure of the ISM toward the Galactic Center. The FUSE data were obtained under program P101. FUSE was operated for NASA by the Department of Physics and Astronomy at the Johns Hopkins University. 
D.K. is supported by an NSF Astronomy and Astrophysics Postdoctoral Fellowship under award AST-2102490. We thank Isabel Rebollido for valuable conversations on the FEROS spectrograph. The FUSE and HST STIS data presented in this paper were obtained from the Mikulski Archive for Space Telescopes (MAST) at the Space Telescope Science Institute. The ESO FEROS spectrum was obtained from the ESO Archive Science Portal.

The FUSE and HST STIS data presented in this paper were obtained from
the Mikulski Archive for Space Telescopes (MAST) at the Space
Telescope Science Institute. The specific observations analyzed
can be accessed via 
\dataset[10.17909/rrnk-3e58]{http://dx.doi.org/10.17909/rrnk-3e58}.
\end{acknowledgments}

%

\vspace{5mm}
\facilities{FUSE, HST(STIS), ESO(FEROS)}


\software{\texttt{astropy} \citep{astropy2018},  
          \textsc{Cloudy} \citep{ferland2017}, 
          \texttt{linetools} \citep{prochaska2017},
          \textsc{VPFIT} \citep{carswell2014},
          \texttt{CalFUSE} \citep{dixon2009},
          \texttt{calSTIS} \citep{dressel2007},
          \texttt{FEROS-DRS} \citep{kaufer1999}
          }


\appendix
\restartappendixnumbering

\section{Global Continuum Fitting}\label{sec:global}

\begin{figure}[!hb]
    \centering
    \includegraphics[width=\textwidth]{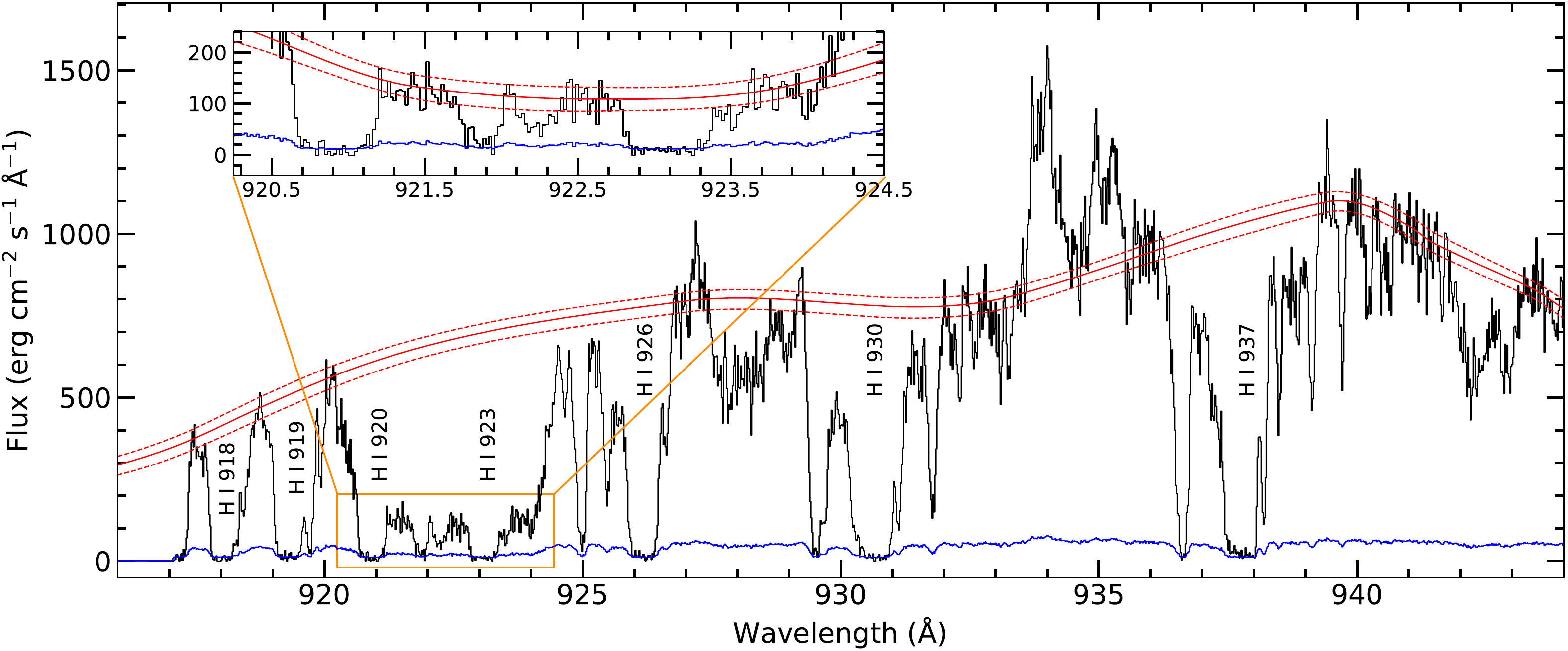}
    \caption{A portion of the FUSE SiC2 spectrum from 916--944 {\AA} in which several higher order \hi\ Lyman series lines are identified. The flux and 1$\sigma$ error in the flux are shown in black and blue, respectively. A global fit to the stellar continuum is traced by a solid red curve, with a high and low stellar continuum placement marked by dashed red lines. The orange box shows a local fit to a depressed portion of the stellar continuum created by stellar line blanketing, from $\sim$920--925 {\AA}.}
    \label{fig:app_ly}
\end{figure}

Stellar absorption lines present a continuum-fitting challenge, particularly in the complex far-UV FUSE spectra used to derive the \hi\ column density in the HVC under study.
Our approach followed in the analysis is to fit the continuum \emph{locally} around each \hi\ Lyman series line of interest, since this allows us to account for stellar absorption lines,
which are present in the continuum when the radiation field encounters the HVC.
However, for completeness here we consider a \emph{global} continuum fitting process, which fits the \hi\ continuum over a larger range (916--944 \AA) in the SiC2 channel. This approach neglects stellar absorption features but ensures the continua are continuous between adjacent lines in the Lyman series. Wavelength regions free from stellar emission and absorption lines defined the flux of the global continuum. 
We show a global fit to the stellar flux in the vicinity of the higher order \hi\ Lyman series lines in Figure \ref{fig:app_ly}, in which areas of emission and absorption due to both stellar and interstellar features can be seen. We performed a simultaneous Voigt profile fit on the same \hi\ lines modeled with the global continua, using the mid-, high-, and low-continuum fits shown in Figure \ref{fig:app_ly}. A separate local continuum was applied to the depressed region resulting from stellar line blanketing and which contains \ion{H}{1} $\lambda$923. Using the global continuum fit (which has a slightly higher continuum placement than the local fits) results in log $N_\mathrm{HI}$ = 15.69$\pm$0.03$\pm$ 0.03 for the HVC, where the first error is the statistical error due to photon noise and the second error is the systematic continuum-placement error. This corresponds to a moderate difference of $\Delta$log $N_\mathrm{HI}$ = 0.15 dex between the \hi\ column densities derived by the global and local continuum methods. 

\section{Decontamination of Molecular Hydrogen}\label{sec:h2-decontaminate}

\begin{figure}
    \centering
    \includegraphics[width=\textwidth]{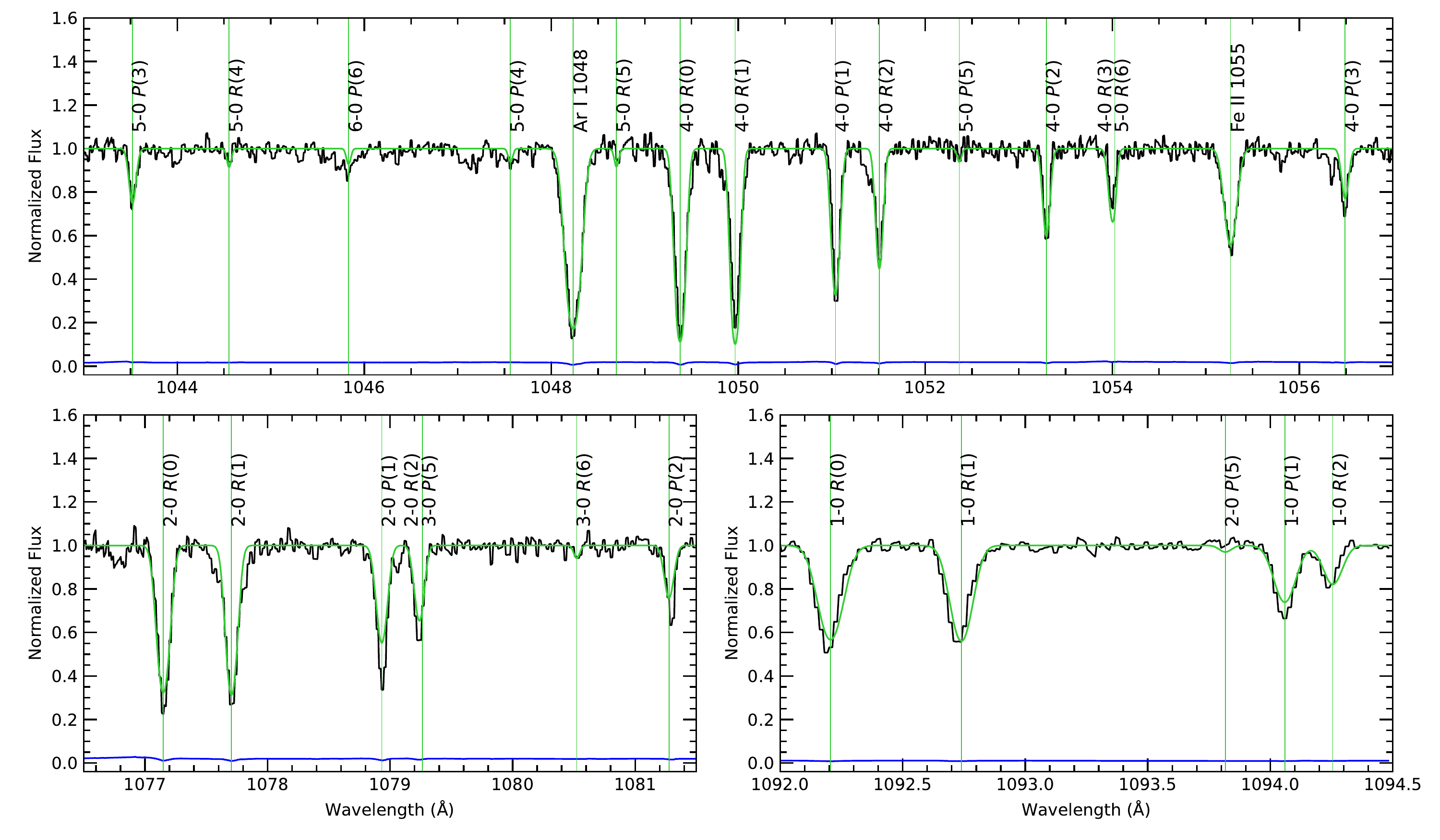}
    \caption{We performed Voigt profile fits to vibrationally and rotationally excited H$_2$ lines from the FUSE LiF1 and LiF2 channels in order to account for molecular blending with \hi. In each panel, the flux and 1$\sigma$ error in the flux are shown in black and blue, respectively. Voigt profile fits to the data are shown in green. Green vertical lines mark the positions of the H$_2$ absorption lines. Top panel: LiF1 channel H$_2$ transitions from upper vibrational levels 4 and 5, spanning rotational levels $J=0-6$. Fits to \ion{Ar}{1} $\lambda$1048 and \ion{Fe}{2} $\lambda$1055 are included. Bottom left panel: LiF1 channel H$_2$ transitions from upper vibrational levels 2 and 3, including rotational levels $J=0,1,2,5,6$. Bottom right panel: LiF2 channel H$_2$ transitions from upper vibrational levels 1 and 2, including rotational levels $J=0,1,2,5$.
    }
    \label{fig:app_mol}
\end{figure}

The FUSE LiF1 and LiF2 channels provide us with an opportunity to model isolated H$_2$ lines. This effort is important because it enables us to account for the Milky Way's molecular contribution to the interstellar absorption lines. In particular, contamination from Milky Way H$_2$ affected the higher order Lyman series \hi\ in the SiC2A channel and \ion{O}{6} in the LiF1 channel. We performed simultaneous single-component Voigt profile fits to multiple H$_2$ transitions from rotational levels $J$ = 0--6, allowing both the velocity and $b$-value to vary freely for the stronger $J$ = 0--4 transitions. The $b$-values of the significantly weaker $J$ = 5,6 transitions were fixed to the $b$-value of the $J$ = 4 transition in order to deter unrealistic column densities for these fainter lines. These results of these fittings are shown in Figure \ref{fig:app_mol}.

\bibliography{ref}{}
\bibliographystyle{aasjournal}



\end{document}